\documentclass[aps,prx,10pt,twocolumn,superscriptaddress]{revtex4}

\usepackage{graphicx}
\usepackage{amssymb, amsmath}
\usepackage{color}
\usepackage{ulem}
\usepackage{bm}
\usepackage{bbm}
\usepackage{graphicx}
\usepackage{subfigure}
\usepackage{dcolumn}
\usepackage{mathtools}
\usepackage{pifont}
\usepackage{diagbox}
\usepackage{braket}
\usepackage{verbatim}
\usepackage{diagbox}
\usepackage{float}

\newfloat{FIGSI}{tb}{si}
\floatname{FIGSI}{FIG. SI}

\def\nn{\nonumber\\}
\def\angs{\mbox{\normalfont\AA}}

\def\I{\uppercase\expandafter{\romannumeral 1}}
\def\II{\uppercase\expandafter{\romannumeral 2}}
\def\III{{\uppercase\expandafter{\romannumeral 3}}}
\def\IV{{\uppercase\expandafter{\romannumeral 4}}}
\def\V{{\uppercase\expandafter{\romannumeral 5}}}
\def\VI{{\uppercase\expandafter{\romannumeral 6}}}
\def\VII{{\uppercase\expandafter{\romannumeral 7}}}

\def\i{\lowercase\expandafter{\romannumeral 1}}
\def\ii{\lowercase\expandafter{\romannumeral 2}}
\def\iii{{\lowercase\expandafter{\romannumeral 3}}}
\def\iv{{\lowercase\expandafter{\romannumeral 4}}}
\def\v{{\lowercase\expandafter{\romannumeral 5}}}
\def\vi{{\lowercase\expandafter{\romannumeral 6}}}
\def\vii{{\lowercase\expandafter{\romannumeral 7}}}

\begin{document}

\title{Generic continuum model formalism for moir\'e superlattice systems}
\author{Bo Xie}
\affiliation{School of Physical Science and Technology, ShanghaiTech Laboratory for Topological Physics, State Key Laboratory of Quantum Functional Materials, ShanghaiTech University, Shanghai 201210, China}%

\author{Jianqi Huang}
\affiliation{Liaoning Academy of Materials, Shenyang 110167, China}

\author{Jianpeng Liu}
\email[]{liujp@shanghaitech.edu.cn}
\affiliation{School of Physical Science and Technology, ShanghaiTech Laboratory for Topological Physics, State Key Laboratory of Quantum Functional Materials, ShanghaiTech University, Shanghai 201210, China}%
\affiliation{Liaoning Academy of Materials, Shenyang 110167, China}

\date{\today}
\begin{abstract}
The moir\'e superlattice system provides an excellent platform for exploring various novel quantum phenomena such as  superconductivity, correlated insulator and the fractional quantum anomalous Hall effects. To theoretically tackle the diverse correlated and topological states emerging from moir\'e superlattices containing thousands of atoms, one usually adopts an effective low-energy continuum model based on which the electron-electron effects are further taken into account. However, the construction of an  accurate continuum model remains a challenging task, particularly for complex moir\'e superlattice systems such as twisted transition metal dichalcogenides (tTMD) systems. 
In this work, we develop a formalism for constructing  generic continuum models that are applicable for arbitrary moir\'e superlattices and are extrapolatable to any twist angles. Our key insight is that the microscopic electronic properties are intrinsic properties of the system, which should remain invariant across all twist angles; the lattice relaxations act as external inputs that vary with twist angles and are coupled with the electrons, and the coupling coefficients are characterized by intrinsic parameters. This partition enables a universal description of how the continuum model evolves with twist angle using a single set of model parameters. In order to extract the model parameters, we design a numerical workflow based on data from first principles density functional theory calculations. As a case study, we apply this framework to twisted bilayer MoTe$_{2}$, where the lattice relaxations are treated utilizing machine learning force fields (MLFFs) methods. Following our workflow, we obtain a single set of  model parameters that accurately reproduce first-principles results for twisted MoTe$_{2}$, including electronic band structures, charge density distributions and Chern numbers, at three different twist angles. Furthermore, the model extrapolates robustly to smaller twist angles. Our work not only provides a more precise understanding of the microscopic properties of moir\'e superlattices, but also lays a foundation for future theoretical studies of low-energy electronic properties in generic  moir\'e superlattice systems. 
\end{abstract}

\maketitle


\section{Introduction}
Moir\'e superlattice systems have aroused  tremendous research interests in recent years. In both graphene-based and transition metal dichalcogenide (TMD)-based moir\'e structures, experiments have uncovered diverse novel phenomena, including unusual superconductivity \cite{cao-nature18-supercond,dean-tbg-science19,marc-tbg-19, efetov-nature19, efetov-nature20, young-tbg-np20, li-tbg-science21, cao-tbg-nematic-science21,KFMak_superconduc_wse2_nature25, dean_superconduc_wse2_nature25}, correlated insulating states \cite{cao-nature18-mott,efetov-nature19,tbg-stm-pasupathy19,tbg-stm-andrei19,tbg-stm-yazdani19, tbg-stm-caltech19, young-tbg-science19, efetov-nature20, young-tbg-np20, li-tbg-science21, yazdani_review_24},  quantum anomalous Hall effects \cite{young-tbg-science19, sharpe-science-19, efetov-arxiv20,yazdani-tbg-chern-arxiv20,efetov-tbg-chern-arxiv20,pablo-tbg-chern-arxiv21, feldman_qah_wse2_science24, KFMak_qshe_tmote2_nature24} and so on. Most recently, fractional quantum anomalous Hall effects have been reported in twisted bilayer MoTe$_{2}$ and rhombohedral-stacked pentalayer and hexalayer graphene \cite{xiaodongxu_fci_tmote2-nature23,shanjie_fci_tmote2-nature23, litinxin_fci_tmote2_prx23, lulong_fci_penta_nature24, xie-nm2025}. A striking feature of these systems is their extreme sensitivity to twist angle. 
An important character of a moir\'e superlattice system is the large moir\'e supercell containing thousands of atoms, with the characteristic moir\'e length scale $\sim 10\,$nm. This structural property enables them to host nearly flat electronic bands in the corresponding mini Brillouin zone, many of which carry non-trivial topological characters due to the vectorial form of the moir\'e potential \cite{macdonald-pnas11,song-tbg-prl19,  yang-tbg-prx19, po-tbg-prb19, origin-magic-angle-prl19, jpliu-prb19}. The interplay between electronic correlations and non-trivial topology is believed to underpin the wealth of phenomena observed in these systems, making them ideal platforms to explore fundamental problems in condensed matter physics \cite{balents-review-tbg,andrei-review-tbg,jpliu-nrp21,kang-tbg-prl19, Uchoa-ferroMott-prl,xie-tbg-2018, wu-chiral-tbg-prb19,  zaletel-tbg-2019, wu-tbg-collective-prl20, zaletel-tbg-hf-prx20,jpliu-tbghf-prb21,zhang-tbghf-arxiv20,hejazi-tbg-hf, kang-tbg-dmrg-prb20,kang-tbg-topomott,yang-tbg-arxiv20,Bernevig-tbg3-arxiv20, Lian-tbg4-arxiv20,regnault-tbg-ed,zaletel-dmrg-prb20,macdonald-tbg-ed-arxiv21, lamparski2020soliton,DaiXi-phonon-2023prb, wu-prl18, lian-prl18, lian-tbg-prl19, sharma-tbg-phonon-nc21, paschen_review_24, wuquansheng_dft_pentalayer_prb24, DasSarma_integer_fractional_pentalayer_peb24, Ashvin_ah_multilayer_graphen1_prb24, Ashvin_ah_multilayer_graphene2_prb24, zhangyahui_fqah_multilayer_graphene_prl24, guozhongqing_fci_prb24, Jennifer_cano_superconduc_twse2_prl25, DasSarma_superconduc_ttmd_prb25}.

To unravel the physics of moir\'e superlattices, researchers have developed a variety of theoretical models, with continuum models standing out as particularly versatile \cite{santos-tbg-prl07, macdonald-pnas11, castro-neto-prb12, koshino-prb17,xie_phonon_prb23,vafek-continuum-prb23}. This is because each moir\'e supercell typically contains thousands of atoms, thus any theoretical studies based on microscopic atomic orbitals would be extremely demanding. Taking twisted bilayer graphene as an example, if only the $p_z$ orbitals of each carbon site are taken into account, there would be about thirteen thousand orbitals at the magic angle, not to mention the spin degeneracy. Any theoretical studies including electron-electron interactions, even at the Hartree-Fock level, would be impractical if based on such a large number of orbitals. Nevertheless, since the essential physical properties are governed by low-energy electrons of the moir\'e length scale, it is unnecessary to include all the microscopic information at the atomistic length scale. Rather, an effective continuum model that accurately captures the low-energy, long wavelength properties would be highly desirable, and would serve as a faithful and feasible starting point to further treat the electron-electron interaction problem.

For twisted bilayer graphene, such a continuum model proposed by Bistrizer and MacDonald not only successfully predicts the existence of flat bands at the magic angle, but also correctly describes the topological characters of the flat bands \cite{macdonald-pnas11}. 
In 2019, the continuum model construction is further extended to transition metal dichalcogenides (TMD) including twisted bilayer MoTe$_2$ and WSe$_2$/MoSe$_2$ heterostructure. The earliest version of the continuum model for moir\'e TMD systems only includes the leading Fourier components of the intralayer and interlayer moir\'e potentials, which are fitted using density functional theory (DFT) calculations of laterally shifted bilayer TMD system without twist \cite{wufengcheng_twsit_tmd_continuum_prl19}. Subsequent extensions incorporate the higher Fourier components of moir\'e potentials, and the parameters are determined by fitting large-scale DFT calculations for the twisted moir\'e system including lattice relaxation effects \cite{fuliang_charge_order_ttmd_prb21, fuliang_ah_fci_ttmd_prb23, zhangyang_transfer_learning_continuum_tmote2_cp24,xiaodi_fci_tmote2_prl24, zhangyang_multiple_chern_band_tmote2_prl25, wuquansheng_dft_tmote2_prb24}. Although these continuum models are  successful in the study of various correlated and topological states emerging from moir\'e TMD systems, they still suffer from critical limitations. Some people attempted to use a single set of model parameters that properly describe the low-energy properties of one twist angle, but such a model can be hardly extrapolated to other twist angles. Some others prefer to fit model parameters separately for each twist angle, but the resulting parameter values vary non-monotonically and even randomly, casting doubt on the physical meaning of those parameters despite good agreement with DFT band structures. 
A more recent approach constructs the continuum model by numerically projecting the DFT Hamiltonian in some localized atomic orbital basis onto the Bloch-sum orbital basis \cite{wuquansheng_without_fit_arxiv25}. Although this method achieves more precise agreement with DFT results, it requires large-scale DFT calculations for every twist angle, leading to a huge cost of computational resources. Additionally, the parameters obtained therein lack clear connection to microscopic physical mechanism. 

In this work, we develop a  theoretical formalism for the construction of continuum models that (in principle) include lattice relaxation effects to arbitrary order, and are applicable for generic moir\'e superlattices and can be extrapolated to any twist angles. Such a formalism overcomes all the aforementioned limitations. Notably, only a single set of universal parameters is introduced to construct the continuum model, which describes the low-energy electronic properties of the system with varying twist angles.
The key insight is to distinguish between two components: (\i) intrinsic properties of the electrons of the parent materials, which are approximately invariant for twist angles, and (\ii) extrinsic lattice relaxation patterns which vary with twist angle and act as external inputs that 
are coupled with the electrons, thus modulate both kinetic energy and moir\'e potential terms of the continuum model. The coupling coefficients to the lattice relaxation fields, however, are still treated as intrinsic parameters which should remain unchanged as the twist angle varies. 
As a result, by explicitly coupling the lattice distortion fields, including both in-plane and out-of-plane components, to the moir\'e potential and kinetic energy, we manage to derive a generic continuum model that is captured by a set of universal parameters that are applicable to all twist angles for a given moir\'e system. 
To extract the model parameters, we design a numerical workflow based on the data of band structures and charge densities calculated from first principles DFT calculations at relatively large twist angles. 
As a case study,  we validate our approach using twisted bilayer MoTe$_{2}$ system. We first perform the lattice relaxation calculations using the machine-learning force fields (MLFF) method, then construct the electronic continuum model that couples to the lattice distortion fields obtained from MLFF. Such a model  is characterized by a set of universal parameters that are applicable to all twist angles. The continuum model constructed by our approach  exhibits excellent agreement with first principles DFT results for band structures, real space distribution of charge densities and topological invariants at various twist angles.  Critically, our model exhibits robust extrapolation to smaller twist angles, eliminating the need for large-scale DFT calculations at small twist angles. 

The paper is organized as follows. In Sec.~\ref{sec:continuum}, we start by introducing the theoretical formalism of constructing the generic continuum model, with elaborated discussions on the effects of lattice distortion fields (or strain fields) on the kinetic energy, interlayer moir\'e potential and intralayer moir\'e potential terms. In Sec.~\ref{sec:numeric} we discuss the numerical workflow to extract the universal parameters of the continuum model, where the DFT data of band structures and band-projected charge densities 
for relatively small moir\'e supercells are used. Such a model then can be applied to various twist angles as long as the corresponding lattice relaxation patterns are given. In Sec.~\ref{sec:app}, we apply our continuum model formalism to twisted bilayer MoTe$_2$, which exhibits excellent agreement with the results obtained from first principles DFT calculations. We further discuss the topological and quantum geometric properties of moir\'e flat bands at a smaller twist angle $3.89^{\circ}$ in this section. In Sec.~\ref{sec:summary} we make a summary of the paper and discuss potential applications of our method.

\section{Generic continuum model}
\label{sec:continuum}
In this section, we present our generic continuum model formalism  for moir\'e superlattice systems. The essential idea underpinning our approach is to  distinguish between the intrinsic properties of electrons and external inputs of the system. Specifically, any moir\'e superlattice system is composed of multiple layered materials, where lattice mismatch and mutual twist give rise to a long-wavelength potential known as the moir\'e potential. Atomic displacements induced by the mutual twist and further lattice distortions would change the lattice potential energy exerted on electrons. From our perspective, the intrinsic properties of moir\'e superlattice systems lie in the microscopic kinetic energy of electrons and the atomic lattice potential exerted on the electrons, the forms of which  rely on the symmetries of the constituent materials, atomic species and atomic orbitals, but should remain invariant across different moir\'e system sizes (twist angle $\theta$). In contrast, the external input variable corresponds to the lattice relaxation patterns, which vary with twist angles, and are coupled to both  the kinetic energy and moir\'e potential terms. The coupling coefficients, nevertheless, are still intrinsic properties of the system and should remain invariant as twist angle varies. In twisted bilayer graphene, such a continuum model can be naturally derived from a microscopic atomistic tight-binding model in the basis of $p_z$-like Wannier orbitals at carbon sites. By virtue of the simple orbital characters of electrons near Dirac point in graphene, the hopping amplitude in this tight-binding model has analytic expression in the Slater-Koster form \cite{moon-tbg-prb13}, from which all the continuum model parameters, including the coupling coefficients to strain fields, can be rigorously derived \cite{koshino-prb17,xie_phonon_prb23}. Thus, the continuum model of twisted bilayer graphene can be considered as a generic one as it can be applied to any twist angle as long as the lattice distortion fields are supplied.
However, obtaining such an accurate and analytic atomistic tight-binding model for arbitrary atom pairs in any material system remains a significant challenge, which hinders the construction of generic continuum models in other moir\'e superlattice systems.  To address this, we propose a semi-analytical continuum model incorporating a single set of universal parameters, which characterize the various kinetic energy, intralayer moir\'e and interlayer moir\'e potential terms that in principle include the couplings to lattice relaxations to arbitrary order. We will derive explicit expressions of the different terms to illustrate how the lattice relaxations couple to low-energy electrons, with  a clear physical meaning assigned to each term. 

\begin{figure}
\includegraphics[width=8cm]{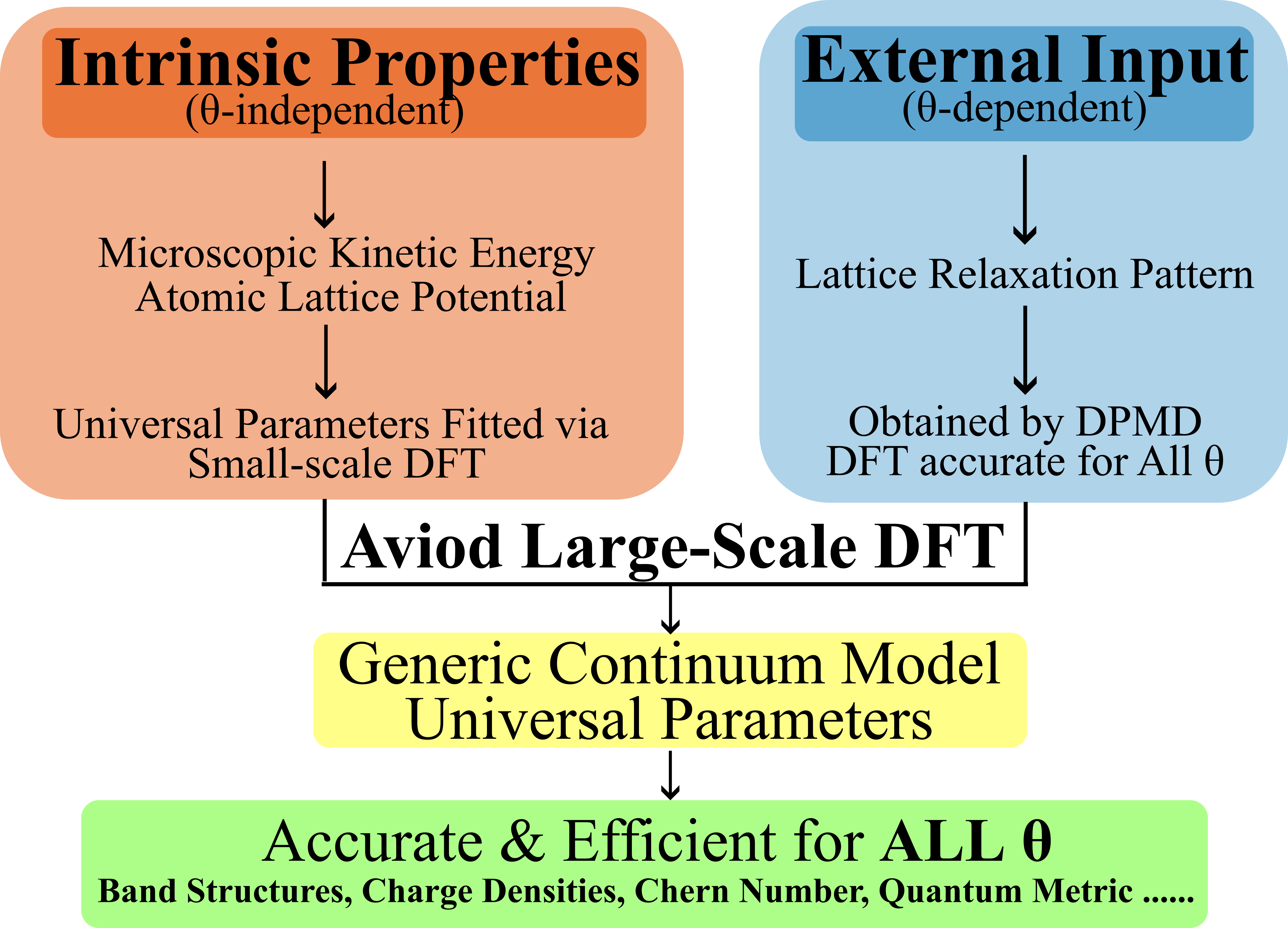}\caption{Schematic illustration of our model, which distinguishes between intrinsic properties of electrons and external inputs. The intrinsic properties correspond to the microscopic kinetic energy of electrons and the atomic lattice potential exerted on the electrons, which remain invariant across different moir\'e system sizes (twist angle $\theta$). They act as one set of universal parameters in our model, fitted using DFT calculations on small-size systems (large $\theta$). The lattice relaxation patterns serve as the twist-angle-dependent external input, modulating the moir\'e potentials and kinetic energy. The lattice relaxation patterns for all twist angle $\theta$ is obtained by Deep Potential Molecular Dynamics (DPMD) simulations. We construct our continuum for all twist angles $\theta$ with one set of universal parameters, which can be extrapolated to small twist angles to study the various physical properties.}
\label{fig1}
\end{figure}

\subsection{General formalism}
We begin by introducing the lattice geometry of moir\'e superlattice systems. Here, we take twisted bilayer system as an example. The lattice vectors of the single layer are $\textbf{a}_{i}^{(l)}$, with corresponding reciprocal lattice vectors $\textbf{a}^{*,(l)}_{i}$, where $l\!=\!1, 2$ labels the layer and $i\!=\!1 ,2$ is the index of the reciprocal vector. Each unit cell contains multiple sublattices, whose positions are given as $\bm{\tau}_{X}^{(l)}$, with $X$ indexing sublattices.
When lattice vectors from two layers are slightly differ with each other, we define moir\'e reciprocal vector as $\textbf{G}^{M}_{i}\!=\!a^{*,(1)}_{i}-a^{*,(2)}_{i}$. The corresponding real space moir\'e superlattice vectors are denoted as $\textbf{L}^{M}_{i}$, which form a commensurate lattice with moir\'e translational symmetry. 
In addition, when two layered materials are stacked together and are twisted by an angle $\theta$, each atom would adjust its position to minimize the total energy. These atomic displacements are coupled to electrons thus modulate the system's low-energy electronic properties. To describe such effects in moir\'e superlattices, we introduce the local atomic shift $\bm{\delta}(\textbf{r})$ at position $\textbf{r}$ within the moir\'e supercell. For an ideal moir\'e superlattice formed by rigidly twisted bilayers (no relaxation), this displacement vector is denoted as $\bm{\delta}_{0}(\textbf{r})$. Atoms may be distorted away from the ideal moir\'e atomic positions. We further decompose such lattice distortions into in-plane and out-of-plane components, denoted as $\textbf{u}^{(l)}_{X}(\textbf{r})$ and $h^{(l)}_{X}(\textbf{r})$, respectively. The total local atomic displacement vector between the two layers at position $\mathbf{r}$ under lattice distortion is given by $\bm{\delta}(\textbf{r})\!=\!\bm{\delta}_{0}(\textbf{r})+\textbf{u}^{(2)}_{X}(\textbf{r})-\textbf{u}^{(1)}_{X}(\textbf{r})+(h^{(2)}_{X}(\textbf{r})-h^{(1)}_{X}(\textbf{r}))\textbf{e}_{z}$. Notably, the length scale of lattice distortion is much larger than the atomic unit cell, allowing us to neglect sublattice indices in the distortion fields, i.e., $\textbf{u}^{(l)}_{X}(\textbf{r})\!=\!\textbf{u}^{(l)}(\textbf{r})$ and $h^{(l)}_{X}(\textbf{r})\!=\!h^{(l)}(\textbf{r})$. We further introduce linear combinations of these distortion fields: $\textbf{u}^{\pm}\!=\!\textbf{u}^{(2)}\pm\textbf{u}^{(1)}$ and $h^{\pm}\!=\!h^{(2)}\pm h^{(1)}$. Here, the minus sign ($-$) describes relative distortions between the two layers, while the plus sign ($+$) corresponds to center-of-mass distortions of the bilayer system as a whole.

We develop a generic continuum model to characterize the low-energy electronic properties of moir\'e superlattice systems.  At the moment, we consider the situation in which the valance band maximum (VBM) or the conduction band minimum (CBM), which dominates the low-energy physics, is located at $\textbf{K}_{l}$ point in the reciprocal space of some hexagonal unit cell. This is the most widely seen situation in moir\'e superlattices such as moir\'e graphene and moir\'e TMD systems. However, it is worth noting that the entire formalism developed in this work is universal, which applies to continuum model expanded around any point (``valley") in reciprocal space. 
The emergence of the low-energy subspace in moir\'e superlattices can be physically interpreted as a two-step process: first, the energy bands are folded in to the mini Brillouin zone of the moir\'e superlattice; second, moir\'e potential modifies these folded low-energy states, leading to the formation of flat bands that are hallmark features of such systems. The basis functions of the continuum model are the Bloch wave functions of the constituent monolayer system near the VBM/CBM:
\begin{equation}
|\Psi_{n,\tilde{\textbf{k}}}\rangle = \sum_{\lambda} C_{\lambda\textbf{G},n,\tilde{\textbf{k}}} |\lambda,  \tilde{\textbf{k}}+\textbf{G} + \textbf{K}_{l} \rangle,
\end{equation}
where $\textbf{G}$ denotes the moir\'e reciprocal lattice vector, $\widetilde{\mathbf{k}}$ is the wavevector within the moir\'e Brillouin zone. $\textbf{K}_{l}$ represents the $\textbf{K}$ point of the $l$th layer after twist, and the index $\lambda$ encapsulates internal degrees of freedom (e.g., spin, sublattice, layer, etc.). In our model, intervalley scattering in the system is neglected such that at the non-interacting level, the continuum model Hamiltonian can be written in a block-diagonal form in valley space. For $\mathbf{K}$ valley, the effective continuum model is formally expressed as:
\begin{align}
\begin{split}
H^{\textbf{K}}=\left[\begin{array}{cc}
				T^{(1)}+V^{(1)} & W \\
				W^{\dagger} & T^{(2)}+V^{(2)}
\end{array}\right].
\end{split}
\end{align}
$T^{(l)}$ and $V^{(l)}$ represents the kinetic energy and intralyer moir\'e potential of layer $l$; while $W$ describes the interlayer moir\'e potential. The essential ideal of our approach is to expand each term as power series of wavevector $\mathbf{k}$ and distortion fields $\{\mathbf{u}^{\pm}, h^{\pm}\}$. 
\begin{align}
&\mathcal{F}_{\lambda,\lambda'}(\textbf{k})=\sum_{\alpha}^{\infty}\sum_{\beta}^{\alpha}\bar{\mathcal{F}}_{\lambda,\lambda'}^{\alpha,\beta} (\{\mathbf{u}^{\pm},h^{\pm}\})\,k_{x}^{\beta}k_{y}^{\alpha-\beta},
\end{align}
where $\mathcal{F}\!=\!\{T^{(l)}, V^{(l)}, W\}$, and $\mathbf{k}=\tilde{\mathbf{k}}+\mathbf{G}+\mathbf{K}_l-\mathbf{K}$ is the wavevector expanded around the untwisted $\mathbf{K}$ point. An alternative way is to expand the kinetic energy and moir\'e potential in polar coordinates of the wavevector $\mathbf{k}=\vert\mathbf{k}\vert(\cos{\phi},\sin{\phi})$:
\begin{align}
&\mathcal{F}_{\lambda,\lambda'}(|\textbf{k}|,\phi)=\sum_{\alpha}^{\infty}\sum_{\beta}^{\alpha}\bar{\mathcal{F}}_{\lambda,\lambda'}^{\alpha,\beta}(\{\mathbf{u}^{\pm},h^{\pm}\})|\textbf{k}|^{\beta}\phi^{\alpha-\beta}.
\end{align}

Specifically, we expand $T^{(l)}$ as a power series of wavevector:
\begin{align}
&T^{(l)}(\tilde{\textbf{k}}+\textbf{G}+\Delta\textbf{K}_{l})_{\lambda,\lambda'}=\nn
&\sum^{\infty}_{\alpha=0}\sum^{\alpha}_{\beta=0}\bar{T}^{(l),\alpha,\beta}_{\lambda,\lambda'}(\tilde{\textbf{k}}+\textbf{G}+\Delta\textbf{K}_{l})_{x}^{\beta}(\tilde{\textbf{k}}+\textbf{G}+\Delta\textbf{K}_{l})_{y}^{\alpha-\beta},
\end{align}
where $\bar{T}^{(l),\alpha,\beta}_{\lambda,\lambda'}$ is independent of momentum, $\Delta\mathbf{K}_l=\mathbf{K}_l-\mathbf{K}$. To incorporate the effects of strain, we further expand $\bar{T}^{(l),\alpha,\beta}_{\lambda,\lambda'}$ as a power series in the strain field:
\begin{align}
&\bar{T}^{(l),\alpha,\beta}_{\lambda,\lambda'}\approx\bar{T}^{(l),\alpha,\beta,0}_{\lambda,\lambda'}+\sum_{s}\sum_{\textbf{G}_{s}}\left[\frac{\partial}{\partial u^{s}_{\textbf{G}_{s}}}\bar{T}^{(l),\alpha,\beta}_{\lambda,\lambda'}\right]S^{s}_{\textbf{G}_{s}}+\nn
&\sum_{s_{1},s_{2}}\sum_{\textbf{G}_{s_1},\textbf{G}_{s_2}}\left[\frac{\partial^{2}}{\partial u^{s_1}_{\textbf{G}_{s_1}}\partial u^{s_2}_{\textbf{G}_{s_2}}}\bar{T}^{(l),\alpha,\beta}_{\lambda,\lambda'}\right]S^{s_{1}}_{\textbf{G}_{s_1}}S^{s_{2}}_{\textbf{G}_{s_2}}+\dots.
\end{align}
In this expansion, $S^{i}_{\textbf{G}_{i}}\!=\!\{\textbf{u}^{\pm}_{\textbf{G}_{i}}, h^{\pm}_{\textbf{G}_{i}}\}$ represents the Fourier components of the strain fields, with $i$ labeling the strain field component. The zeroth-order term $\bar{T}^{(l),\alpha,\beta,0}_{\lambda,\lambda'}$ corresponds to the unstrained system, while higher-order terms capture linear and nonlinear responses of the kinetic energy to strain.

The intralayer moir\'e potential can be expressed as:
\begin{align}
&V^{(l)}(\textbf{r},\textbf{r}')=\sum_{j}V^{(l)}(\tilde{\textbf{k}},\textbf{Q}_{j})_{\lambda,\lambda'}\nn
=&\sum_{j}\sum^{\infty}_{\alpha=0}\sum^{\alpha}_{\beta=0}\bar{V}^{(l),\alpha,\beta}_{\lambda,\lambda'}(\textbf{Q}_{j})\cdot(\tilde{\textbf{k}}+\textbf{G})_{x}^{\beta}(\tilde{\textbf{k}}+\textbf{G})_{y}^{\alpha-\beta}\delta_{\textbf{G}',\textbf{G}+\textbf{Q}_{j}},
\end{align}
where $\bar{V}^{(l),\alpha,\beta}_{\lambda,\lambda'}(\textbf{Q}_{j})$ is independent of $\textbf{k}$. To address the dependency of intralayer interactions to structural perturbations, we further expand $\bar{V}$ with respect to the strain fields:
\begin{align}
\scalebox{0.85}{
$\begin{aligned}
&\bar{V}^{(l),\alpha,\beta}_{\lambda,\lambda'}(\textbf{Q}_{j})\approx \bar{V}^{(l),\alpha,\beta,0}_{\lambda,\lambda'}(\textbf{Q}_{j})+\sum_{s}\sum_{\textbf{G}_{s}}\left[\frac{\partial}{\partial S^{s}_{\textbf{G}_{s}}}\bar{V}^{(l),\alpha,\beta}_{\lambda,\lambda'}(\textbf{Q}_{j})\right]S_{\textbf{G}_{s}}^{s}+\\
&\sum_{s_{1},s_{2}}\sum_{\textbf{G}_{s_{1}},\textbf{G}_{s_{2}}}\left[\frac{\partial^{2}}{\partial S^{s_{1}}_{\textbf{G}_{s_{1}}}\partial S^{s_{2}}_{\textbf{G}_{s_{2}}}}\bar{V}^{(l),\alpha,\beta}_{\lambda,\lambda'}(\textbf{Q}_{j})\right]S_{\textbf{G}_{s_{1}}}^{s_{1}}S_{\textbf{G}_{s_{2}}}^{s_{2}}+\dots.
\end{aligned}$%
}
\label{eq:V-expansion}
\end{align}
This expansion captures how strain modulates interlayer distances and in-plane distortion fields. $\bar{V}^{\alpha,\beta,0}_{\lambda,\lambda'}(\textbf{Q}_{j})$ represents the unstrained reference potential, while the subsequent terms describe strain-induced corrections. $\mathbf{Q}_j$  denotes the moir\'e reciprocal vector. 

We now turn to the interlayer moir\'e potential that arises from electron's hopping between the two layers, the amplitude of which is modulated with moir\'e periodicity.
\begin{align}
&\sum_{j}W(\tilde{\textbf{k}},\textbf{Q}_{j})_{\lambda,\lambda'}\nn
=&\sum_{j}\sum^{\infty}_{\alpha=0}\sum^{\alpha}_{\beta=0}\bar{W}^{\alpha,\beta}_{\lambda,\lambda'}(\textbf{Q}_{j})\cdot(\tilde{\textbf{k}}+\textbf{G})_{x}^{\beta}(\tilde{\textbf{k}}+\textbf{G})_{y}^{\alpha-\beta},\;
\label{eq:w_expansion_k}
\end{align}
where $\bar{W}^{\alpha,\beta}_{\lambda,\lambda'}(\textbf{Q}_{j})$ is independent of $\textbf{k}$. We can further expand $\bar{W}$ in terms of the strain fields:
\begin{align}
\scalebox{0.85}{
$\begin{aligned}
&\bar{W}^{\alpha,\beta}_{\lambda,\lambda'}(\textbf{Q}_{j})\approx \bar{W}^{\alpha,\beta,0}_{\lambda,\lambda'}(\textbf{Q}_{j})+\sum_{s}\sum_{\textbf{G}_{s}}\left[\frac{\partial}{\partial u^{s}_{\textbf{G}_{s}}}\bar{W}^{\alpha,\beta}_{\lambda,\lambda'}(\textbf{Q}_{j})\right]S_{\textbf{G}_{s}}^{s}+\\
&\sum_{s_{1},s_{2}}\sum_{\textbf{G}_{s_{1}},\textbf{G}_{s_{2}}}\left[\frac{\partial^{2}}{\partial S^{s_{1}}_{\textbf{G}_{s_{1}}}\partial S^{s_{2}}_{\textbf{G}_{s_{2}}}}\bar{W}^{\alpha,\beta}_{\lambda,\lambda'}(\textbf{Q}_{j})\right]S_{\textbf{G}_{s_{1}}}^{s_{1}}S_{\textbf{G}_{s_{2}}}^{s_{2}}+\dots.
\end{aligned}$%
}
\label{eq:w_expansion}
\end{align}
This expansion captures the linear and nonlinear responses of the interlayer  potential to lattice relaxations. In the following subsection, we will elaborate on the physical mechanisms underlying this expansion, including how strain modulates interlayer distances and atomic orbital overlaps to alter the interlayer potential.

Symmetries of the moir\'e systems would impose constraints to the allowed terms in the continuum model. Consider a symmetry operation $g$ which is represented by a  matrix $O_g$ in the flavor space. It would imposes the following constraints to the continuum model:
\begin{align}
&\sum^{\infty}_{\alpha=0}\sum^{\alpha}_{\beta=0} \mathcal{F}^{\alpha,\beta}_{\lambda,\lambda'}\,k_{x}^{\beta}\,k_{y}^{\alpha-\beta}\;\nn
=&\sum^{\infty}_{\alpha=0}\sum^{\alpha}_{\beta=0}\sum_{\lambda_{1},\lambda'_{1}}\left[O_{g}\right]^{*}_{\lambda'_{1},\lambda'}\mathcal{F}^{\alpha,\beta}_{\lambda_{1},\lambda'_{1}}\left[O_{g}\right]_{\lambda_{1},\lambda}\,(g^{-1}\mathbf{k})_{x}^{\beta}\,(g^{-1}\mathbf{k})_{y}^{\alpha-\beta}\;.
\end{align}
Again, $\mathcal{F}=\{T, V, W\}$.  $\mathcal{F}_{\lambda,\lambda'}^{\alpha,\beta}$ can be further expanded in power series of strain fields $\mathbf{S}=\{\mathbf{u}^{\pm},h^{\pm}\}$, and symmetry operations would impose constraints on the expansion coefficients. More details  are given in Appendix.

In our model, we distinguish between momentum-independent and momentum-dependent components of the moir\'e potentials: the former describes a purely local potential, while the latter accounts for non-local effects arising from projections to a given subspace. The nonlocalness in real space implies that the Fourier components of the potential depend not only on the moir\'e reciprocal vector $\mathbf{Q}_j$, but also on the incident wavevector $\mathbf{k}$ of the Bloch states. Here, we expand the nonlocal moir\'e potential in polar coordinate of $\mathbf{k}$, which takes the form:
\begin{align}
V_{\mathbf{Q}_j}(\textbf{k})=\sum_{n,m} V^{m,n}_{\mathbf{Q}_j}\,\mathcal{T}_{n}(|\textbf{k}|L_{s}\eta) e^{i m\phi},
\label{eq:v-nonlocal}
\end{align}
where $\mathcal{T}_{n}$ is the $n$-th Chebyshev polynomial. The factor $\eta$ is an arbitrary scaling factor to be determined. We provide the detailed derivation of nonlocal moir\'e potential in Appendix C.

\subsection{Correction to kinetic energy induced by strain}
We evaluate the correction to kinetic energy induced by strain fields, focusing on the microscopic mechanisms that link lattice distortions to changes in kinetic energy. Within the framework of tight-binding theory, we consider a unit cell containing $N_{W}$ Wannier functions. Then we define the displacement vector from $n$-th Wannier function in the ``home" cell to the neighboring $m$-th Wannier function in the $j$th unitcell $\textbf{r}_{nm}^{j}$, which in turn determines the hopping amplitude. Due to the lattice relaxation effects, the atoms would alter their position to minimize the total energy of the system (considering zero temperature). We introduce the three dimensional lattice strain tensor $\hat{\mathbf{u}}$. Under the lattice distortion operation, the displacement vectors vary from $\textbf{r}_{nm}^{j}$ to $(\textbf{r}_{nm}^{j})'\!=\!(1+\hat{\mathbf{u}})\textbf{r}_{nm}^j$. This change in displacement vector leads to a shift in the hopping amplitude given by $\delta t^{n,m}(\textbf{r})\!=\!t^{n,m}(\textbf{r}')-t^{n,m}(\textbf{r})$. The tight binding Hamiltonian is given by: $t^{n,m}(\textbf{k})\!=\!\sum_{j}t^{n,m}e^{i\,\textbf{k}_a\cdot\textbf{r}_{nm}^j}$, where $\textbf{k}_a$ is the wavevector in reciprocal space measured with respect to the atomic Brillouin zone center. We expand the tight binding model around the band edge wavevector $\textbf{K}$ by setting $\textbf{k}_a=\textbf{K}+\textbf{k}$, where $\textbf{k}$ measures the deviation from the band edge.
\begin{align}
&t^{n,m}(\textbf{k}+\textbf{K})\nn
=&\sum_{j}(t_{0}^{n,m}(\textbf{r}_{nm}^j)+\delta t^{n,m}(\textbf{r}_{nm}^j))e^{i(\textbf{K}+\textbf{k})\cdot\textbf{r}_{nm}^j}\nn
\approx&\sum_{j}\,e^{i\textbf{K}\cdot\textbf{r}_{nm}^j}\Big(\,t^{n,m}_{0}(\textbf{r}_{nm}^j)+\delta t^{n,m}(\textbf{r}_{nm}^j)(1+i\textbf{k}\cdot\textbf{r}_{nm}^j+\dots)\,\Big).
\label{eq:strainkinetic}
\end{align}
Here, the first term corresponds to the unstrained Hamiltonian at the band edge, while the subsequent terms describe strain-induced corrections, with the expansion in $\textbf{k}$ capturing the low-energy dispersion. Besides, if the system possesses certain symmetry, it would impose constraints on the expansion parameters (See Appendix) \cite{koshino-tbg-prb17, xie_phonon_prb23}.

\subsection{Correction to intralayer moir\'e potential induced by strain}
We first consider the case of local moir\'e potential. 
Within the intralayer Hamiltonian, the presence of a twisted adjacent layer induces a position-dependent potential $V^{(l)}(\bm{\delta}(\textbf{r}))$ on the $l$-th layer. We model $V^{(l)}(\bm{\delta}(\textbf{r}))$ as a function of local displacement $\bm{\delta}(\textbf{r})$. The  lattice relaxations would alter the relative positions of the atoms from different layers and further affect the potential, which can be formally expressed as $V^{(l)}(\bm{\delta}(\textbf{r}))\!=\!V^{(l)}(\bm{\delta}_{\parallel}(\textbf{r}),h^{-}(\textbf{r}))$, where $\bm{\delta}_{\parallel}(\textbf{r})$  denotes in-plane distortion field and $h^{-}(\mathbf{r})$ denotes the relative out-of-plane distortion field. To explicitly characterize the distortion-induced correction to the intralayer moir\'e potential, we perform a Fourier expansion of $V^{(l)}(\bm{\delta}_{\parallel}(\textbf{r}),h^{-}(\textbf{r}))$ in terms of the in-plane distortion fields.
\allowdisplaybreaks
\begin{align}
\label{eq:vs}
\scalebox{0.9}{
$\begin{aligned}
&V^{(l)}(\bm{\delta}_{\parallel}(\textbf{r}),h^{-}(\textbf{r}))\\
=&\sum_{j}V^{(l)}_{\textbf{G}_j}(h^{-}(\textbf{r}))e^{i\textbf{a}_{j}^{*}\cdot\bm{\delta}_{\parallel}(\textbf{r})}\\
=&\sum_{j}V^{(l)}_{\textbf{G}_j}(h^{-}(\textbf{r}))e^{i\textbf{a}_{j}^{*}\cdot(\bm{\delta}_{0}(\textbf{r})+\textbf{u}^{-}(\textbf{r}))}\\
=&\sum_{j}V^{(l)}_{\textbf{G}_j}(h^{-}(\textbf{r}))e^{i\textbf{G}_{j}\cdot\textbf{r}}\prod_{\textbf{G}^{u}_{m}}\sum_{n_{m}}\frac{1}{n_{m}!}(i\textbf{a}_{j}^{*}\cdot\textbf{u}^{-}_{\textbf{G}^{u}_{m}})^{n_{m}}e^{i n_{m}\textbf{G}^{u}_{m}\cdot\textbf{r}}\\
=&\sum_{j}\sum_{n_{m_{1}},\dots}V^{(l)}_{\textbf{G}_j}(h^{-}(\textbf{r}))\frac{\left(i\textbf{a}^{*}_{j}\cdot\textbf{u}^{-}_{\textbf{G}^{u}_{m_{1}}}\right)^{n_{m_{1}}}}{n_{m_{1}}!}\frac{\left(i\textbf{a}^{*}_{j}\cdot\textbf{u}^{-}_{\textbf{G}^{u}_{m_{2}}}\right)^{n_{m_{2}}}}{n_{m_{2}}!}\dots\\
&\ \ \ \ \ \dots e^{i\textbf{G}_{j}\cdot\textbf{r}}e^{i(n_{m_1}\textbf{G}^{u}_{m_1}+n_{m_2}\textbf{G}^{u}_{m_2}+\dots)\cdot\textbf{r}}.
\end{aligned}$%
}
\end{align}
The product over $\textbf{G}^{u}_{m}$ and sum over $n_{m}$ account of the contribution of multiple strain harmonics, with $n_{m}$ account for the order of expansion in the strain fields. The expansion explicitly demonstrates how strain fields modified the intralyer moir\'e potential. It directly corresponds to the general form introduced in Eq.~\eqref{eq:V-expansion}. The non-relaxed (bare) moir\'e potential is recovered by setting all expansion orders to zero ($n_{m_{i}}=0$).  The  dependence of intralayer moir\'e potential on the relative out-of-plane distortion $h^{-}(\mathbf{r})$ can be treated in a similar manner. However, since it is expected that the out-of-plane corrugation has a much weaker effect on the intralayer potential than on the interlayer one, we assume that the intralayer potential only depends on the average interlayer distance $\langle h^{-}(\mathbf{r})\rangle=d_0$. More discussions on this point are given in Sec.~\ref{sec:pro_mo}. 

\subsection{Correction to interlayer moir\'e potential induced by strain}
Again, we first focus on the local interlayer moir\'e potential by restricting to the $\mathbf{k}$-independent term of the potential, i.e. the $\alpha\!=\!0, \beta\!=\!0$ term in Eq.~\eqref{eq:w_expansion_k}.  Here we provide an explicit expression of the interlayer moir\'e potential. We begin by considering the microscopic hopping process between atomic sites in adjacent layers. Under a two-particle approximation, the hopping amplitude between a site in one layer and another site in the adjacent layer is denoted as $t(\textbf{r}+d\textbf{e}_{z})$. Although the exact functional form of $t$ depends on the specific material systems, we can always define the two dimensional (2D) Fourier transformation of the real-space hopping:
\begin{equation}
W(\textbf{q})=\frac{1}{S_{0}d_{0}}\int \mathrm{d}^{3}r\, t(\textbf{r}+d\textbf{e}_{z})e^{-i\textbf{q}\cdot\textbf{r}},
\end{equation}
where $S_{0}$ is the area of the unit cell and $d_{0}$ is the average interlayer distance. We consider both in-plane distortion $\textbf{u}^{(l)}(\textbf{r})$ and out-of-plane distortion $h^{(l)}(\textbf{r})$  for the $l$-th layer. The Bloch state in layer $l$ with two-dimensional wave vector $\textbf{k}$ is defined as:
$\ket{\textbf{k},l}=1/\sqrt{N}\sum_{\textbf{R}^{(l)}}e^{i\textbf{k}\cdot(\textbf{R}^{(l)}+\bm{\tau}_{X})}\ket{\textbf{R}^{(l)}_{X}+\textbf{u}^{(l)}+h^{(l)}\textbf{e}_{z}}$. The interlayer hopping matrix element is then given by:
\begin{align}
\scalebox{0.85}{
$\begin{aligned}
&\bra{\textbf{k}',l'}U\ket{\textbf{k},l}\nn
=&\frac{1}{N}\frac{S_{0}d_{0}}{(2\pi)^{3}}\int \mathrm{d}^{3}p\ W(\textbf{p})\sum_{\textbf{R}\in\textbf{R}^{(1)}_{X}}e^{i(\textbf{k}-\textbf{p}_{\parallel})\cdot(\textbf{R}_{\parallel}+\bm{\tau}_{X})-i\textbf{p}_{\parallel}\cdot\textbf{u}^{(1)}-i\ p_{z}(\tau_{z}+h^{(1)})}\nn
&\times\sum_{\textbf{R}'\in\textbf{R}^{(2)}_{X'}} e^{-i(\textbf{k}'-\textbf{p}_{\parallel})\cdot(\textbf{R}_{\parallel}'+\bm{\tau}_{X'})+i\textbf{p}_{\parallel}\cdot\textbf{u}^{(2)}+ip_{z}(\tau_{X'}+h^{(2)})}\;,
\end{aligned}$%
}
\end{align}
where the three dimensional wavevector $\mathbf{p}$ is decomposed into the in-plane component $\mathbf{p}_{\parallel}$ and out-of-plane component $p_z$.
The lattice distortion fields can be Fourier expanded as: $\textbf{u}^{(l)}(\textbf{R})\!=\!\sum_{\textbf{G}^{u}_{m}}\textbf{u}^{(l)}_{\textbf{G}^{u}_{m}}e^{i\textbf{G}^{u}_{m}\cdot\textbf{R}}$ and $h^{(l)}(\textbf{R})\!=\!\sum_{\textbf{G}^{h}_{m}}h^{(l)}_{\textbf{G}^{h}_{m}}e^{i\textbf{G}^{h}_{m}\cdot\textbf{R}}$.
We utilize the identity: $\sum_{\textbf{R}\in\textbf{R}^{(l)}}e^{i\textbf{q}\cdot\textbf{R}}\!=\!N\sum_{\textbf{g}}\delta_{\textbf{p}_{\parallel},\textbf{g}^{(l)}}$, where $\mathbf{g}$ denotes reciprocal vector of the atomic lattice of $l$th layer. Then, the interlayer hopping term is expressed as:
\begin{align}
\scalebox{0.85}{
$\begin{aligned}
&\bra{\textbf{k}',l'}U\ket{\textbf{k},l}=\sum_{\textbf{g}^{(1)},\textbf{g}^{(2)}}\sum_{n_{1},\dots}\sum_{n_{h,1},\dots}\sum_{n_{1}'\dots}\sum_{n'_{h,1}\dots}e^{-i(\textbf{g}^{(1)}\cdot\bm{\tau}_{X}-\textbf{g}^{(2)}\cdot\bm{\tau}_{X'})}\nn
&\ \ \ \ \ \ \ \ \ \gamma(\textbf{Q})\delta_{\textbf{k}+\textbf{g}^{(1)}+n_{1}\textbf{q}_{1}+n_{h,1}\textbf{q}_{h,1}+\dots,\textbf{k}'+\textbf{g}^{(2)}-n_{1}\textbf{q}_{1}-n_{h,1}\textbf{q}_{h,1}+\dots},
\end{aligned}$%
}
\end{align}
where $\textbf{Q}=\textbf{Q}_{\parallel}+p_{z}\textbf{e}_{z}$, and the in-plane wavevector
\begin{align}
&\textbf{Q}_{\parallel}=\textbf{k}+\textbf{g}^{(1)}+n_{1}\textbf{G}^{u}_{1}+n_{h,1}\textbf{G}^{h}_{1}+n_{2}\textbf{G}^{u}_{2}+n_{h,2}\textbf{G}^{h}_{2}+\dots\nn
&=\textbf{k}'+\textbf{g}^{(2)}-n'_{1}\textbf{G}^{u}_{1}-n'_{h,1}\textbf{G}^{h}_{1}-n'_{2}\textbf{G}^{u}_{2}-n'_{h,2}\textbf{G}^{h}_{2}+\dots,
\end{align}
where $\{\textbf{G}^{u}_{1},\textbf{G}^{u}_{2},\textbf{G}^{h}_{1},\textbf{G}^{h}_{2},\dots\}$ are moir\'e reciprocal vectors. The effective interlayer hopping amplitude is expressed as:
\begin{align}
&\gamma(\textbf{Q}_{\parallel}+p_{z}\textbf{e}_{z})\nn
\approx&\frac{d_{0}}{2\pi}\int\mathrm{d}p_{z}\,W(\textbf{Q}_{\parallel}+p_{z}\textbf{e}_{z})e^{i\,p_{z}d_{0}}\nn
&\times\sum_{n_1,n_{h,1},\dots}\frac{\left[i\textbf{Q}_{\parallel}\cdot\textbf{u}^{-}_{\textbf{G}^{u}_{1}}\right]^{n_{1}+n'_{1}}}{n_{1}!n'_{1}!}\frac{\left[i\textbf{Q}_{\parallel}\cdot\textbf{u}^{-}_{\textbf{G}^{u}_{2}}\right]^{n_{2}+n'_{2}}}{n_{2}!n'_{2}!}\dots\nn
&\times\frac{\left[i\,p_{z}h^{-}_{\textbf{G}^{h}_{1}}\right]^{n_{h,1}+n'_{h,1}}}{n_{h,1}!n'_{h,1}!}\frac{\left[i\,p_{z}h^{-}_{\textbf{G}^{h}_{2}}\right]^{n_{h,2}+n'_{h,2}}}{n_{h,2}!n'_{h,2}!}\dots
\end{align}
 The interlayer hopping matrix element can be expressed in real-space representation as
\begin{equation}
\bra{\textbf{k}',l'}U\ket{\textbf{k},l}=\frac{1}{S}\int \mathrm{d}^{2}r e^{i(\textbf{k}-\textbf{k}')\cdot\textbf{r}}U(\textbf{r}),
\end{equation}
where the moir\'e potential with relaxed lattice structure is given by:
\begin{align}
\scalebox{0.85}{
$\begin{aligned}
U(\textbf{r})=\sum_{j}\frac{d_{0}}{2\pi}\int^{\infty}_{-\infty}\mathrm{d} p_{z}e^{i(\textbf{Q}_{j}\cdot\textbf{u}^{-}(\textbf{r})+\textbf{G}_{j}\cdot\textbf{r}+p_{z}(d_{0}+h^{-}(\textbf{r})))}W(\textbf{Q}_{j}+p_{z}\textbf{e}_{z})
\end{aligned}$%
}
\end{align}

The integration over $p_{z}$ can be carried out using the trick of integration by part \cite{xie_phonon_prb23}. We first consider the case without out-of-plane distortion, i.e., the terms with $n_{h,1}+n'_{h,1}=0$, $n_{h,2}+n'_{h,2}=0$, $\dots$. The integration over $p_{z}$ leads to
\begin{align}
\scalebox{0.9}{
$\begin{aligned}
&\frac{d_{0}}{2\pi}\int \mathrm{d} p_{z}e^{i p_{z}d_{0}}W(\textbf{Q}_{j}+p_{z}\textbf{e}_{z})=\frac{1}{S_{0}}\int \mathrm{d}^{2}r\,t(\textbf{r}+d_{0}\textbf{e}_{z})e^{-i\textbf{Q}_{j}\cdot\textbf{r}}.
\end{aligned}$%
}
\end{align}
This term is denoted as $\tilde{W}_{0}$. This corresponds to the unstrained interlayer hopping amplitude, denoted as $\tilde{T}_{0}$ (with unit of eV). For the term that is expanded to the first order of out-of-plane strain ($n_{h,1}+n'_{h,1}=1, n_{h,2}+n'_{h,2}=0, \dots$), the integral becomes:
\allowdisplaybreaks
\begin{align}
&\frac{d_{0}}{2\pi}\int\mathrm{d}p_{z}\,W(\textbf{Q}_{\parallel}+p_{z}\textbf{e}_{z})e^{i\,p_{z}d_{0}}\times\left[i\,p_{z}\right]\nonumber\\
=&\frac{1}{S_{0}}\int\mathrm{d}^{2}r\left.\frac{\mathrm{d}}{\mathrm{d}d'}t(\textbf{r}+d'\textbf{e}_{z})\right|_{d'=d_{0}}e^{-i\textbf{Q}\cdot\textbf{r}}.
\end{align}
This term, denoted as $\tilde{W}_{1}$ (with unit of eV/$\angs$), describes the linear dependency of interlayer moir\'e potential to out-of-plane distortions. Higher-order terms, such as those proportional to second-order out-of-plane strain ($n_{h,1}+n'_{h,1}=1$, $n_{h,2}+n'_{h,2}=1$, and $n_{h,2}+n'_{h,2}=0, \dots$),yield:
\begin{align}
&-\frac{d_{0}}{2\pi}\int \mathrm{d} p_{z}e^{i p_{z}d_{0}}W(\textbf{Q}_{j}+p_{z}\textbf{e}_{z})\times i p_{z}h^{-}_{\textbf{G}^{h}_{m_1}}i p_{z}h^{-}_{\textbf{G}^{h}_{m_2}}\nn
=&-\frac{1}{S_{0}}\int \mathrm{d}^{2}r \left.\frac{\mathrm{d}^{2}}{\mathrm{d}^{2} d'}t(\textbf{r}+d'\textbf{e}_{z})\right|_{d'=d_{0}}e^{-i\textbf{Q}_{j}\cdot\textbf{r}}h^{-}_{\textbf{G}^{h}_{m_1}}h^{-}_{\textbf{G}^{h}_{m_2}}\;,
\end{align}
which is denoted as $\tilde{W}_{2}$ with unit of eV/$\angs^{2}$. Combining these contributions, the interlayer hopping amplitude is expressed as:
\begin{align}
\label{eq:ws}
\scalebox{0.8}{
$\begin{aligned}
&\gamma(\textbf{Q}_{\parallel}+p_{z}\textbf{e}_{z})\\
=&\sum_{n_{m_1},n_{m_1}',\dots}\,\Big(\,\tilde{W}_{0}(d_{0})\frac{\left[i\textbf{Q}_{\parallel}\cdot\textbf{u}^{-}_{\textbf{G}_{m_1}}\right]^{n_{m_1}+n'_{m_1}}}{n_{m_1}!n'_{m_1}!}\frac{\left[i\textbf{Q}_{\parallel}\cdot\textbf{u}^{-}_{\textbf{G}_{m_2}}\right]^{n_{m_2}+n'_{m_2}}}{n_{m_2}!n'_{m_2}!}\dots\\
&+\tilde{W}_{1}(d_{0})h_{\textbf{G}^{h}_{1}}^{-}\frac{\left[i\textbf{Q}_{\parallel}\cdot\textbf{u}^{-}_{\textbf{G}_{m_1}}\right]^{n_{m_1}+n'_{m_1}}}{n_{m_1}!n'_{m_1}!}\frac{\left[i\textbf{Q}_{\parallel}\cdot\textbf{u}^{-}_{\textbf{G}_{m_2}}\right]^{n_{m_2}+n'_{m_2}}}{n_{m_2}!n'_{m_2}!}\dots\\
&+\tilde{W}_{2}(d_{0})h_{\textbf{G}^{h}_{1}}^{-}h_{\textbf{G}^{h}_{2}}^{-}\frac{\left[i\textbf{Q}_{\parallel}\cdot\textbf{u}^{-}_{\textbf{G}_{m_1}}\right]^{n_{m_1}+n'_{m_1}}}{n_{m_1}!n'_{m_1}!}\frac{\left[i\textbf{Q}_{\parallel}\cdot\textbf{u}^{-}_{\textbf{G}_{m_2}}\right]^{n_{m_2}+n'_{m_2}}}{n_{m_2}!n'_{m_2}!}\dots\,\Big).
\end{aligned}$%
}
\end{align}
This expansion explicitly realizes the strain-dependent interlayer moir\'e potential described in Eq.~\eqref{eq:w_expansion}, with each term quantifying how lattice distortions, both in-plane and out-of-plane, modify the hopping amplitude. The lattice distortion would further scatters states from two layers and modulate the hopping amplitudes. 

\subsection{Properties of moir\'e potentials}
\label{sec:pro_mo}
In this section, we discuss the key properties of the moir\'e potentials. A unifying theme in our framework is that all long wavelength moir\'e potentials originate from microscopic atomic hopping processes and atomic lattice potentials, which serve as the intrinsic properties underlying the emergent superlattice behavior.

For the intralayer moir\'e potential, it mostly originates from the charge modulation of the adjacent twisted layer. This potential is inherently sensitive to structural distortions: in-plane displacements alter the lateral alignment of charge distributions, while out-of-plane corrugations modify the vertical distance between charges of the two layers. To capture this, we model the intralayer potential as a function of both in-plane and out-of-plane distortions, denoted $V^{(l)}(\bm{\delta}_{\parallel}(\textbf{r}),h^{-}(\textbf{r}))$. Since the moir\'e potential is more sensitive to large-scale average spacing than local ripples, here for the intralayer moir\'e potential, we neglect the spatial variation in the out-of-plane direction, i.e., $h^{-}(\textbf{r}\rightarrow d_{0})$. Here $d_{0}$ is the average interlayer distance. When the average layer spacing deviates from $d_{0}$ by a small amount $\delta d$, say, due to the change of twist angle, the intralyer potential undergos a corresponding change:
\begin{align}
V^{(l)}_{\textbf{G}}(d_{0}+\delta d)\approx& V^{(l)}_{\textbf{G}}(d_{0})+\left.\frac{\partial V^{(l)}_{\textbf{G}}}{\partial d'}\right|_{d'=d_{0}}\delta d +\dots\nn
=&V^{(l)}_{\textbf{G}}(d_{0})+ V^{(l),1}_{\textbf{G}}(d_{0})\delta d+\dots.
\label{eq:vd}
\end{align}
Here, $V^{(l),1}_{\textbf{G}}(d_{0})\!=\!\partial V^{(l)}_{\textbf{G}}/\partial \left. d'\right|_{d'=d_{0}}$ quantifies the linear dependence of the intralyer moir\'e potential to the average interlayer distance.

The interlayer moir\'e potential  directly inherits its properties from the microscopic interlayer hopping $t$. This amplitude exhibits distinct contributions from structural distortions. $\tilde{W}_{0}(d_{0})$ describes the moir\'e potential induced by two layers without corrugation effects. $\tilde{W}_{1}(d_{0})$ and $\tilde{W}_{2}(d_{0})$ describe the first order and second order corrections of corrugation effects to the moir\'e potential. Notably, the amplitude of $\tilde{W}_{0}$, $\tilde{W}_{1}$ and $\tilde{W}_{2}$ are fixed for given $d_{0}$, as they depend only on the microscopic hopping $t$ and the average interlayer distance. A small shift $\delta d$ from a reference $d_{0}$ modifies $\tilde{W}_{0}$ through a Taylor expansion:
\begin{align}
&\tilde{W}_{0}(d_{0}+\delta d)\nn
\approx&\tilde{W}_{0}(d_{0})+\left.\frac{\partial\tilde{W}}{\partial d'}\right|_{d'=d_{0}}\delta d+\frac{1}{2}\left.\frac{\partial^{2}\tilde{W}}{\partial d'^{2}}\right|_{d'=d_{0}}\delta d^{2}+\dots\nn
=&\tilde{W}_{0}(d_{0})+\tilde{W}_{1}(d_{0})\delta d+\frac{1}{2}\tilde{W}_{2}(d_{0})\delta d^{2}+\dots.
\label{eq:td}
\end{align}
This expansion reveals a dual role for $\tilde{W}_{1}$ and $\tilde{W}_{2}$: they not only describe the correction of corrugation effects to the interlayer moir\'e potential, but also describe the variation of interlayer hopping parameters at different $d_{0}$. 

The average interlayer distance $d_{0}$ is not a fixed quantity but is determined by the lattice relaxations of the moir\'e superlattice system, which is the external input in our framework.  As the system size (twist angle) changes, the stacking registry of the layers is modified, altering the average interlayer distance $d_{0}$. This, in turn, builds up the twist angle dependent moir\'e potential.


\section{Numerical workflow}
\label{sec:numeric}

After the above formulation, we come up with a semi-analytical continuum model with several parameters to be fixed, denoted as $\mathcal{P}\!=\!$ $\{V^{(l)}_{\textbf{G}}$, $V^{(l),1}_{\textbf{G}}$, $\tilde{W}_{0}$, $\tilde{W}_{1}$, $\tilde{W}_{2},\dots\}$. From Eq.~\eqref{eq:vs} and Eq.~\eqref{eq:ws}, we can determine $V^{(l)}_{\textbf{G}}$ and $\tilde{W}_{0}$ from the DFT results with unrelaxed lattice structures. $V^{(l)}_{\textbf{G}}$, $\tilde{W}_{0}$, $\tilde{W}_{1}$ and $\tilde{W}_{2}$ are included to capture the influence induced by lattice relaxation, based on the DFT results with fully relaxed lattice structures. Besides, $V^{(l),1}_{\textbf{G}}$, $\tilde{W}_{1}$ and $\tilde{W}_{2}$ encode the dependence of moir\'e potentials on the average interlayer distance $d_{0}$ according to Eq.~\eqref{eq:vd} and Eq.~\eqref{eq:td}.  We can describe systems with different twist angles using one single set of  consistent parameters. As a result, we partition $\mathcal{P}$ into two parts, $\mathcal{P}_{u}\!=\!\{V^{(l)}_{\textbf{G}},\tilde{W}_{0}\}$ and $\mathcal{P}_{d}\!=\!\{V^{(l),1}_{\textbf{G}},\tilde{W}_{1}, \tilde{W}_{2},\dots\}$. Here, $\mathcal{P}_{u}$ stands for the zeroth-order derivative of the moir\'e potential with respect to average interlayer distance $d_{0}$, which can be fitted from DFT results with unrelaxed lattice structures. It is worth noting that although the parameters $\mathcal{P}_u$ are determined from DFT data without lattice relaxations, they can also determine the couplings of electrons to in-plane lattice distortions as shown in Eqs.~\eqref{eq:w-strain}-\eqref{eq:v-strain}.
 $\mathcal{P}_{d}$ encodes both the correction induced by corrugation effects and the variation of moir\'e potentials due to the change of average interlayer distance $d_{0}$, which can be fitted from DFT results with fully relaxed lattice structures.

\begin{figure}
\includegraphics[width=8cm]{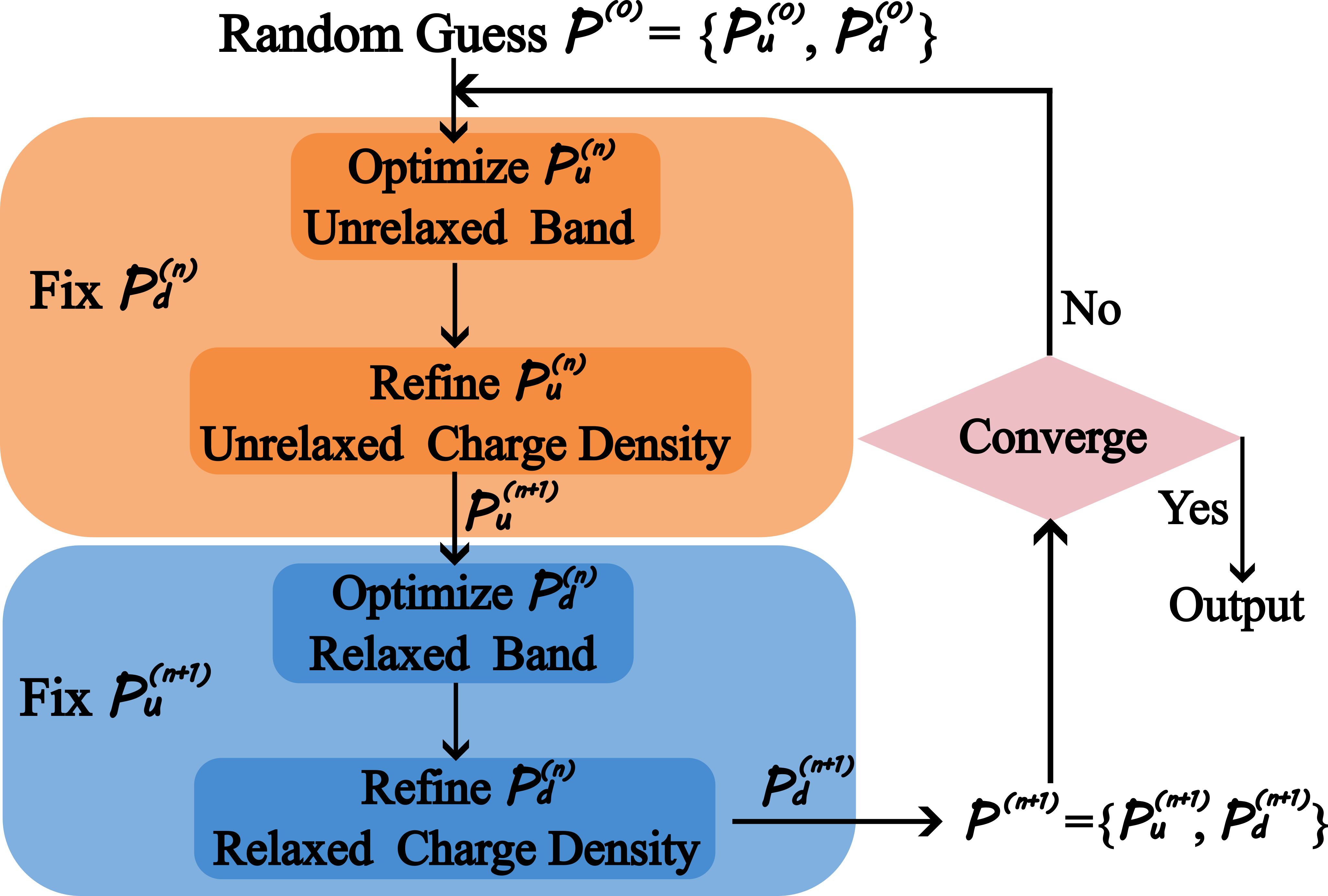}
\caption{Schematic of the numerical workflow for solving the model parameters. The parameter set $\mathcal{P}$ is partitioned into two subsets: $\mathcal{P}_{u}$  (zeroth-order derivative of the moir\'e potential with respect to average interlayer distance $d_{0}$) and $\mathcal{P}_{d}$ (both the correction induced by corrugation effects and the variation of moir\'e potentials due to the change of average interlayer distance $d_{0}$). Starting from a random initial guess for $\mathcal{P}$, we iteratively optimize $\mathcal{P}_{u}$ and $\mathcal{P}_{d}$  in an alternating manner until convergence is achieved.}
\label{fig2}
\end{figure}

The goal of optimization is to minimize the discrepancy between the continuum model’s predictions and the reference data from first-principles DFT calculations. We define a loss function $L$ that quantifies this mismatch through several key physical quantities: band structures, real space distribution of charge densities, and Chern number. Specifically, we define the loss function as follows:
\begin{equation}
L=\sum_{\textbf{k},n,i}(f^{i}_{n,\textbf{k}}-g^{i}_{n,\textbf{k}})^{2}w^{i}_{n},
\end{equation}
where $f$ represents the physical quantity calculated by continuum model, $g$ represents the physical quantity calculated by DFT. $n$ represents the sampling points of the system. $i$ represents the index of moir\'e system (differentiated by superlattice size or twist angle). $w^{i}_{n}$ are assigned weights. The total loss is the sum of contributions from all quantities across all the target twist angles. However, optimizing all parameters in $\mathcal{P}$ simultaneously risks trapping in local minima. In order to avoid this risk, we design an iterative, stepwise workflow that alternates between $\mathcal{P}_{u}$ and $\mathcal{P}_{d}$. The process is as follows (schematically illustrated in Fig.~\ref{fig2}):


$\expandafter{\romannumeral 1}$) Initialize parameters: Start with a random initial guess $\mathcal{P}^{(0)}\!=\!\{\mathcal{P}_{u}^{(0)},\mathcal{P}_{d}^{(0)}\}$.

$\expandafter{\romannumeral 2}$) Optimize $\mathcal{P}_{u}^{(0)}$: Keep $\mathcal{P}_{d}^{(0)}$ fixed, and minimize the loss function $L$ for the unrelaxed band structure for all three systems (with three different twist angles) to fit $\mathcal{P}_{u}^{(0)}$. Then, refine $\mathcal{P}_{u}^{(0)}$ using  the DFT data of unrelaxed charge density. Updates $\mathcal{P}_{u}^{(0)}$ to $\mathcal{P}_{u}^{(1)}$.

$\expandafter{\romannumeral 3}$) Optimize $\mathcal{P}_{d}^{(0)}$: Fix $\mathcal{P}_{u}^{(1)}$ and minimize the loss function $L$ for relaxed band structure for all three systems ( with three different twist angles) to fit $\mathcal{P}_{d}^{(0)}$. Then refine $\mathcal{P}_{d}^{(0)}$ based using the DFT data of relaxed charge density. Updates $\mathcal{P}_{d}^{(0)}$ to $\mathcal{P}_{d}^{(1)}$.

$\expandafter{\romannumeral 4}$) Check convergence: $\mathcal{P}^{(1)}\!=\!\{\mathcal{P}_{u}^{(1)},\mathcal{P}_{d}^{(1)}\}$. If the change in parameters falls below a threshold, the process converges. Otherwise, repeat steps $\expandafter{\romannumeral 2}$) - $\expandafter{\romannumeral 3}$).

\section{Application to twisted bilayer MoTe$_2$}
\label{sec:app}
As an application of our theoretical framework, we focus on the twisted bilayer MoTe$_{2}$ system, which exhibits rich correlated and topological physics such as fractional quantum anomalous Hall effects. Single layer MoTe$_{2}$ crystallizes in a hexagonal lattice, with a band structure characterized by a valence band maximum (VBM) located at $\textbf{K}/\mathbf{K}'$ point \cite{yaowang_spin-valley_lock_tmd_prl12}. We begin with a bilayer MoTe$_{2}$ system, where two layers are aligned without any twist. Upon introducing a small twist angle $\theta$ between the layers, the periodicity of the system is extended, forming a moir\'e superlattice. The low-energy bands around $\mathbf{K}$ and $\mathbf{K}'$ points would be folded into moir\'e Brillouin zone and renormalized by the moir\'e potentials. 

\subsection{Lattice relaxation patterns}
To accurately capture the lattice relaxation effects in twisted bilayer MoTe$_{2}$, we perform a molecular dynamics calculation based on the machine learning force fields (MLFFs). This approach combines the accuracy of DFT methods with the computational efficiency needed to simulate large moir\'e superlattices, making it an ideal method for studying lattice relaxation effects in moir\'e superlattice systems. 

The MLFFs are built using Deep Potential Molecular Dynamics (DeepMD-kit) framework \cite{dpmd, dpmd2}, which employs deep neural networks to learn the atomic force field from DFT training data. Our training data consists of 4000 untwisted $3\times3$ bilayer MoTe$_{2}$ configurations with random in-plane shifts, out-of-plane shifts and atomic positional perturbations. For each configuration, we compute the total energy, force and virial matrix using the Vienna Ab Initio Simulation Package (VASP) package \cite{vasp} with the Perdew-Burke-Ernzerhof (PBE) \cite{PBE} exchange-correlation functional and projector-augmented wave (PAW) pseudopotentials \cite{PAW}. For van der Waals corrections, we use the density dependent screened Coulomb (dDsC) dispersion correction (IVDW=4 in VASP) \cite{ivdw4_1, ivdw4_2}. Using the trained MLFFs, we perform large-scale lattice relaxation simulations using Large-scale Atomic/Molecular Massively Parallel Simulator (LAMMPS) package \cite{lammps}. 

\begin{figure}
\includegraphics[width=8cm]{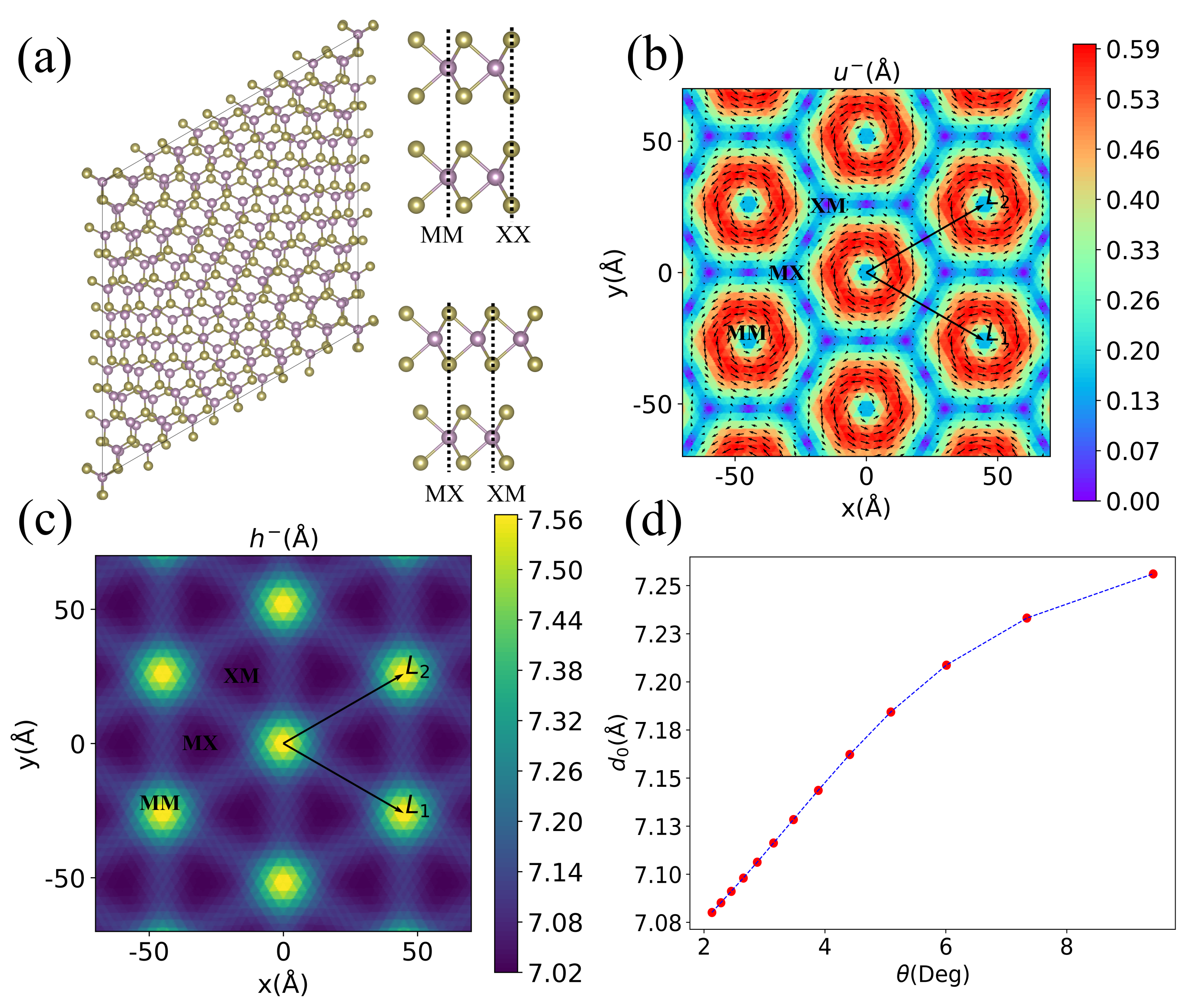}
\caption{Lattice relaxation patterns of twisted bilayer MoTe$_{2}$. (a) Moir\'e  lattice structure of twisted bilayer MoTe$_{2}$ with $\theta\!=\!9.43^{\circ}$. (b) The real space distribution of the in-plane relative distortions at $\theta\!=\!3.89^{\circ}$. The colorbar represents the amplitudes of distortions. The arrows represent the directions of the distortion fields. (c) The real space distribution of the out-of-plane relative distortions (the interlayer distance) at $\theta\!=\!3.89^{\circ}$. The colorbar represents the value of the interlayer distance. (d) The average interlayer distance ($d_{0}$, in units of $\angs$) as a function of twist angle.}
\label{fig3}
\end{figure}

In Fig.~\ref{fig3}, we present the lattice relaxation patterns for twisted bilayer MoTe$_{2}$. In Fig.~\ref{fig3}(b), we present the real space distribution of the in-plane relative lattice relaxation pattern $\textbf{u}^{-}(\textbf{r})$ of twisted MoTe$_{2}$ with $\theta\!=\!3.89^{\circ}$. The colorbar encodes the amplitudes of in-plane distortions and the arrows indicate the directions of in-plane distortions. The in-plane relaxation field forms a rotational pattern around the $MM$ point, where Mo atoms from the two layers are aligned. The maximum amplitude of in-plane distortion is $0.59\,\angs$. Fig.~\ref{fig3}(c) describes the real space distribution of the relative out-of-plane distortion $h^{-}(\textbf{r})$, which corresponds to the real-space distribution of interlayer distance. The interlayer distance exhibits a maximum  of $7.56\,\angs$ at the $MM$ point and a minimum of $7.02\,\angs$ at $MX/XM$ point. In Fig.~\ref{fig3}(d), we present the average interlayer distance $d_{0}$ as a function of twist angle $\theta$. $d_{0}$ decreases monotonically from $7.25\,\angs$ at $\theta\!=\!9.43^{\circ}$ to $7.08\,\angs$ at $\theta\!=\!2.13^{\circ}$. As we mentioned before, this variation in $d_{0}$ directly modulates the moir\'e potential. Our results show consistency with previous studies \cite{wuquansheng_dft_tmote2_prb24, zhangyang_transfer_learning_continuum_tmote2_cp24}.

These relaxation patterns serve as critical inputs to our continuum model, enabling it to capture the strain-induced modifications to the electronic structure of twisted MoTe$_{2}$. By combining these structural data with our parameterized moir\'e potentials, we can systematically investigate how twist angle and lattice relaxation together shape the low-energy electronic properties.

\subsection{Electronic continuum model}
Here we introduce the effective continuum model describing the low-energy electronic properties in twisted MoTe$_{2}$. This system preserves $C_{3z}$, $C_{2y}$ and time-reversal $\mathcal{T}$ symmetries. In particular, $C_{3z}$ and $C_{2y}\mathcal{T}$ are symmetries of the continuum model Hamiltonian of one valley, while $C_{2y}$ flips the two valleys. These symmetries impose stringent constraints on the form of the Hamiltonian and its parameters, reducing the degrees of freedom and ensuring physical consistency.

We consider one Wannier function localized within a unit cell of single layer MoTe$_{2}$, which is constructed from the Bloch functions of the highest valence band. The hopping amplitude between two Wannier functions, separated by a vector $\textbf{r}+d\textbf{e}_{z}$, is expressed as:
\begin{align}
&t(\textbf{r}+d\textbf{e}_{z})\nn
=&t_{0}(\textbf{r}+d\textbf{e}_{z})+t_{3}(\textbf{r}+d\textbf{e}_{z})(\cos(3\theta_{12}(\textbf{r}))+\cos(3\theta_{21}(\textbf{r})))
\end{align}
Here, $t_{0}$ describes the isotropic hopping and $t_{3}$ captures the anisotropic component with $C_{3}$ rotational symmetry. The angles $\theta_{12}$/$\theta_{21}$ denote the orientation of the hopping vector relative to the primary bond directions in the first/second layer. Our focus on low-energy physics restricts attention to states near the VBM at $\textbf{K}$ and $\textbf{K}'$ points in the atomic Brillouin zone.

The kinetic energy for each layer $l$ is modeled as:
\begin{align}
T^{(l)}=\frac{\hbar^{2}(\textbf{k}-\textbf{K}^{(l)})^{2}}{2m^{*}}+V_{s},
\end{align}
where $m^{*}\!=\!0.62\,m_{e}$ is the effective mass of electrons near the VBM in single layer MoTe$_{2}$. $V_{s}$ accounts for the strain induced correction to the kinetic energy. The symmetries impose constraints on Eq.~\eqref{eq:strainkinetic} (See Appendix), resulting in
\begin{align}
V_{s}=\sum_{\textbf{G}_{s}}i\left[\begin{matrix}
G_{s, x} & G_{s, y}
\end{matrix}
\right]\left[\begin{matrix}
u^{(l)}_{\textbf{G}_{s},x}\\
u^{(l)}_{\textbf{G}_{s},y} 
\end{matrix}
\right]A^{(l)}e^{i\textbf{G}_{s}\cdot\textbf{r}}.
\label{kinetic-strain}
\end{align}

For the interlayer moir\'e potential, we define the three first-neighbor wavevectors as  $\textbf{Q}^{\mu}_{[0], 1}=\textbf{K}^{\mu}$, $\textbf{Q}^{\mu}_{[0], 2}=\textbf{K}^{\mu}+\mu\textbf{a}^{*}_{1}$ and $\textbf{Q}^{\mu}_{[0], 3}=\textbf{K}^{\mu}+\mu(\textbf{a}^{*}_{1}+\textbf{a}^{*}_{2})$, where $\mu=\pm$ is the valley index. Similarly, we consider the moir\'e potential contributed by the second and third nearest neighbor interlayer hopping in momentum space. The corresponding wavevector $\textbf{Q}^{Ref}$ is defined as:  $\textbf{Q}^{\mu}_{[1],1}=\textbf{K}^{\mu}-\mu\textbf{a}^{*}_{2}$, $\textbf{Q}^{\mu}_{[1],2}=\textbf{K}^{\mu}+\mu\textbf{a}^{*}_{2}$ and $\textbf{Q}^{\mu}_{[1],3}=\textbf{K}^{\mu}+\mu(2\textbf{a}^{*}_{1}+\textbf{a}^{*}_{2})$ and $\textbf{Q}^{\mu}_{[2],1}\!=\!\textbf{K}^{\mu}-\mu(\textbf{a}^{*}_{1}+\textbf{a}^{*}_{2})$, $\textbf{Q}^{\mu}_{[2],2}\!=\!\textbf{K}^{\mu}+\mu(\textbf{a}^{*}_{1}+2\textbf{a}^{*}_{2})$, $\textbf{Q}^{\mu}_{[2],3}\!=\!\textbf{K}^{\mu}+\mu(\textbf{a}^{*}_{1}-\textbf{a}^{*}_{2})$, $\textbf{Q}^{\mu}_{[2],4}\!=\!\textbf{K}^{\mu}-\mu\textbf{a}^{*}_{1}$, $\textbf{Q}^{\mu}_{[2],5}\!=\!\textbf{K}^{\mu}+2\mu(\textbf{a}^{*}_{1}+\textbf{a}^{*}_{2})$, $\textbf{Q}^{\mu}_{[2],6}\!=\!\textbf{K}^{\mu}+2\mu\textbf{a}^{*}_{1}$. .

The interlayer hopping matrix element can be expressed as
\begin{align}
&\bra{\textbf{k}',l'}U\ket{\textbf{k},l}=\sum_{i,j}\sum_{n_{1},\dots}\sum_{n_{h,1},\dots}\sum_{n_{1}'\dots}\sum_{n'_{h,1}\dots}\nn
&\gamma(\textbf{Q}_{[i],j}+p_{z}\textbf{e}_{z})\delta_{\textbf{k}',\textbf{k}+\textbf{G}_{[i],j}+(n_{1}+n'_{1})\textbf{G}^{u}_{1}+(n_{h,1}+n'_{h,1})\textbf{G}^{h}_{1}+\dots},
\end{align}
where $\textbf{G}_{[0],1}=0$, $\textbf{G}_{[0],2}=\mu\textbf{G}^{M}_{1}$ and $\textbf{G}_{[0],3}=\mu(\textbf{G}^{M}_{1}+\textbf{G}^{M}_{2})$, $\textbf{G}^{M}_{1}$ and $\textbf{G}^{M}_{2}$ are the moir\'e reciprocal lattice vectors.

Following the derivation in Sec.~\ref{sec:continuum}, the effective hopping can be expressed as an integral over the out-of-plane momentum, which yields meaningful hopping parameters. For the nearest neighbor hopping:
\begin{align}
&\frac{d_{0}}{2\pi}\int \mathrm{d} p_{z}e^{i p_{z}d_{0}}W(\textbf{Q}_{j}+p_{z}\textbf{e}_{z})\nn
=&\frac{1}{S_{0}}\int \mathrm{d}^{2}r\,t(\textbf{r}+d_{0}\textbf{e}_{z})e^{-i\textbf{Q}_{[0],j}\cdot\textbf{r}}=\tilde{W}_{[0],0}.
\end{align}
We have a complex number $\tilde{W}_{[0],0}$ in unit of eV. Due to the gauge freedom, we set $\tilde{W}_{[0],0}$ as a real number. Similarly, we have $\tilde{W}_{[0],1}$ and $\tilde{W}_{[0],2}$. We also consider the second harmonic terms of the interlayer hopping. The corresponding momentum transfer are defined as: $\textbf{G}_{[1],1}=-\mu\textbf{G}^{M}_{2}$, $\textbf{G}_{[1],2}=\mu\textbf{G}^{M}_{2}$ and $\textbf{G}_{[1],3}=\mu(2\textbf{G}^{M}_{1}+\textbf{G}^{M}_{2})$. For the third harmonic terms, we have: $\textbf{G}_{[2],1}=-\mu(\textbf{G}^{M}_{1}+\textbf{G}^{M}_{2})$, $\textbf{G}_{[2],2}=\mu(\textbf{G}^{M}_{1}+2\textbf{G}^{M}_{2})$, $\textbf{G}_{[2],3}=\mu(\textbf{G}^{M}_{1}-\textbf{G}^{M}_{2})$, $\textbf{G}_{[2],4}=-2\mu\textbf{G}^{M}_{1}$, $\textbf{G}_{[2],5}=2\mu(\textbf{G}^{M}_{1}+\textbf{G}^{M}_{2})$ and $\textbf{G}_{[2],6}=2\mu\textbf{G}^{M}_{1}$. The interlayer moir\'e potential is:
\begin{align}
\scalebox{0.8}{
$\begin{aligned}
&\gamma(\textbf{Q}_{[i]}+p_{z}\textbf{e}_{z})\\
&=\sum_{n_{m_1},n_{m_1}', \dots}\,\Big(\,\tilde{W}_{[i],0}\times\frac{\left[i\textbf{Q}_{[i]}\cdot\textbf{u}^{-}_{\textbf{G}_{m_1}}\right]^{n_{m_1}+n'_{m_1}}}{n_{m_1}!n'_{m_1}!}\frac{\left[i\textbf{Q}_{[i]}\cdot\textbf{u}^{-}_{\textbf{G}_{m_2}}\right]^{n_{m_2}+n'_{m_2}}}{n_{m_2}!n'_{m_2}!}\dots\\
&+\tilde{W}_{[i],1}h_{\textbf{G}^{h}_{1}}^{-}\times\frac{\left[i\textbf{Q}_{[i]}\cdot\textbf{u}^{-}_{\textbf{G}_{m_1}}\right]^{n_{m_1}+n'_{m_1}}}{n_{m_1}!n'_{m_1}!}\frac{\left[i\textbf{Q}_{[i]}\cdot\textbf{u}^{-}_{\textbf{G}_{m_2}}\right]^{n_{m_2}+n'_{m_2}}}{n_{m_2}!n'_{m_2}!}\dots\\
&+\tilde{W}_{[i],2}h_{\textbf{G}^{h}_{1}}^{-}h_{\textbf{G}^{h}_{2}}^{-}\times\frac{\left[i\textbf{Q}_{[i]}\cdot\textbf{u}^{-}_{\textbf{G}_{m_1}}\right]^{n_{m_1}+n'_{m_1}}}{n_{m_1}!n'_{m_1}!}\frac{\left[i\textbf{Q}_{[i]}\cdot\textbf{u}^{-}_{\textbf{G}_{m_2}}\right]^{n_{m_2}+n'_{m_2}}}{n_{m_2}!n'_{m_2}!}\dots\,\Big).
\end{aligned}$
}
\label{eq:w-strain}
\end{align}
Here $\tilde{W}_{[i],0}$ describe the interlayer moir\'e potential of the rigid moir\'e superlattice without lattice relaxations. $\tilde{W}_{[i],1}$ and $\tilde{W}_{[i],2}$ (with units eV/$\angs$ and ev/$\angs^{2}$, respectively) describe the first and second corrections to the interlayer hopping induced by out-of-plane corrugations. The index $[i]$ denotes the $(i+1)$-th Fourier component. This expression explicitly describes how to couple the lattice relaxation fields to electronic properties.

To capture the essential features of the intralyer moir\'e potential, we include contributions from the first, second, and third Fourier components in our model:
\allowdisplaybreaks
\begin{align}
&V^{(l)}(\bm{\delta}_{\parallel}(\textbf{r}),d_{0})\nn
=&\sum_{i,j}\sum_{n_{m_{1}},n_{m_{1}}', \dots}\,V^{(l)}_{\textbf{G}_{[i],j}}\frac{\left(i\textbf{a}^{*}_{j}\cdot\textbf{u}^{-}_{\textbf{G}^{u}_{m_{1}}}\right)^{n_{m_{1}}}}{n_{m_{1}}!}\frac{\left(i\textbf{a}^{*}_{j}\cdot\textbf{u}^{-}_{\textbf{G}^{u}_{m_{2}}}\right)^{n_{m_{2}}}}{n_{m_{2}}!}\dots\nn
&\dots e^{i\textbf{G}_{[i],j}\cdot\textbf{r}}e^{i(n_{m_1}\textbf{G}^{u}_{m_1}+n_{m_2}\textbf{G}^{u}_{m_2}+\dots)\cdot\textbf{r}}.
\label{eq:v-strain}
\end{align}
Here, $V^{(l)}_{\textbf{G}_{[i],j}}$ denotes the amplitude of the $(i+1)$-th Fourier component. The product terms account for the corrections induced by in-plane lattice distortions.
Due to the $C_{3}$ rotational symmetry, $V^{(l)}_{\textbf{G}_{[i],j}}=V^{(l)}_{C_{3}(\textbf{G}_{[i],j})}$. The intralayer moir\'e potentials from two different layers within the same valley are linked through $C_{2y}\mathcal{T}$ symmetry operation. Thus, $V^{(1)}_{\textbf{G}_{j}}=V^{(2)}_{-\textbf{G}_{j}}$.

Here we focus on the local moir\'e potential, neglecting the non-local terms. This approximation is justified by numerical checks (See Appendix), where we find that non-local contributions to the low-energy band structure are small. Nevertheless, our formalism can certainly take into account the effects of nonlocal moir\'e potential, which in Fourier space is expressed as a potential that depends both on the incident wavevector $\mathbf{k}$ and the transferred wavevector $\mathbf{Q}$ due to the potential scattering as expressed in Eq.~\eqref{eq:v-nonlocal}.

Combining the strain-corrected kinetic energy and moir\'e potential, the total Hamiltonian of twisted MoTe$_{2}$ is given by:
\begin{align}
\scalebox{0.80}{
$\begin{aligned}
\begin{split}
H=\left[\begin{array}{cc}
				\frac{\hbar^{2}(\textbf{k}-\textbf{K}_{+})^{2}}{2m^{*}}+V^{(1)}(\textbf{r})+V_{s}^{(1)} & W \\
				W^{\dagger} &  \frac{\hbar^{2}(\textbf{k}-\textbf{K}_{-})^{2}}{2m^{*}}+V^{(2)}(\textbf{r})+V_{s}^{(2)}+\Delta
\end{array}\right].
\end{split}
\end{aligned}$
}
\end{align}
The Hamiltonian is fully characterized by a set of physically meaningful parameters, all derived from microscopic hopping amplitude of some presumable atomistic tight-binding model. And, these parameters are determined by fitting to DFT data: $\{\tilde{W}_{[0],0}$, $\tilde{W}_{[0],1}$, $\tilde{W}_{[0],2}$, $\tilde{W}_{[1],0}$, $\tilde{W}_{[1],1}$, $\tilde{W}_{[2],0}\}$ account for the interlayer moir\'e potential and the corresponding corrections induced by lattice relaxations; $\{V_{\textbf{G}_{[0]}}$, $V_{\textbf{G}_{[0]}}^{1}$, $V_{\textbf{G}_{[1]}}$, $V_{\textbf{G}_{[2]}}\}$ represent the intralayer moir\'e potential and the corresponding lattice-relaxation corrections; The parameter set $\{A\}$, introduced in Eq.~\eqref{kinetic-strain}, denotes the coefficients of the strain-induced corrections to kinetic energy. $\Delta$ represents the interlayer potential drop induced by vertical displacement fields. These parameters encapsulate the system’s response to external input (the lattice relaxations etc.), enabling quantitative predictions of low-energy electronic properties at various twist angles using \textit{a single set of universal parameters}.

\begin{figure*}
\includegraphics[width=16cm]{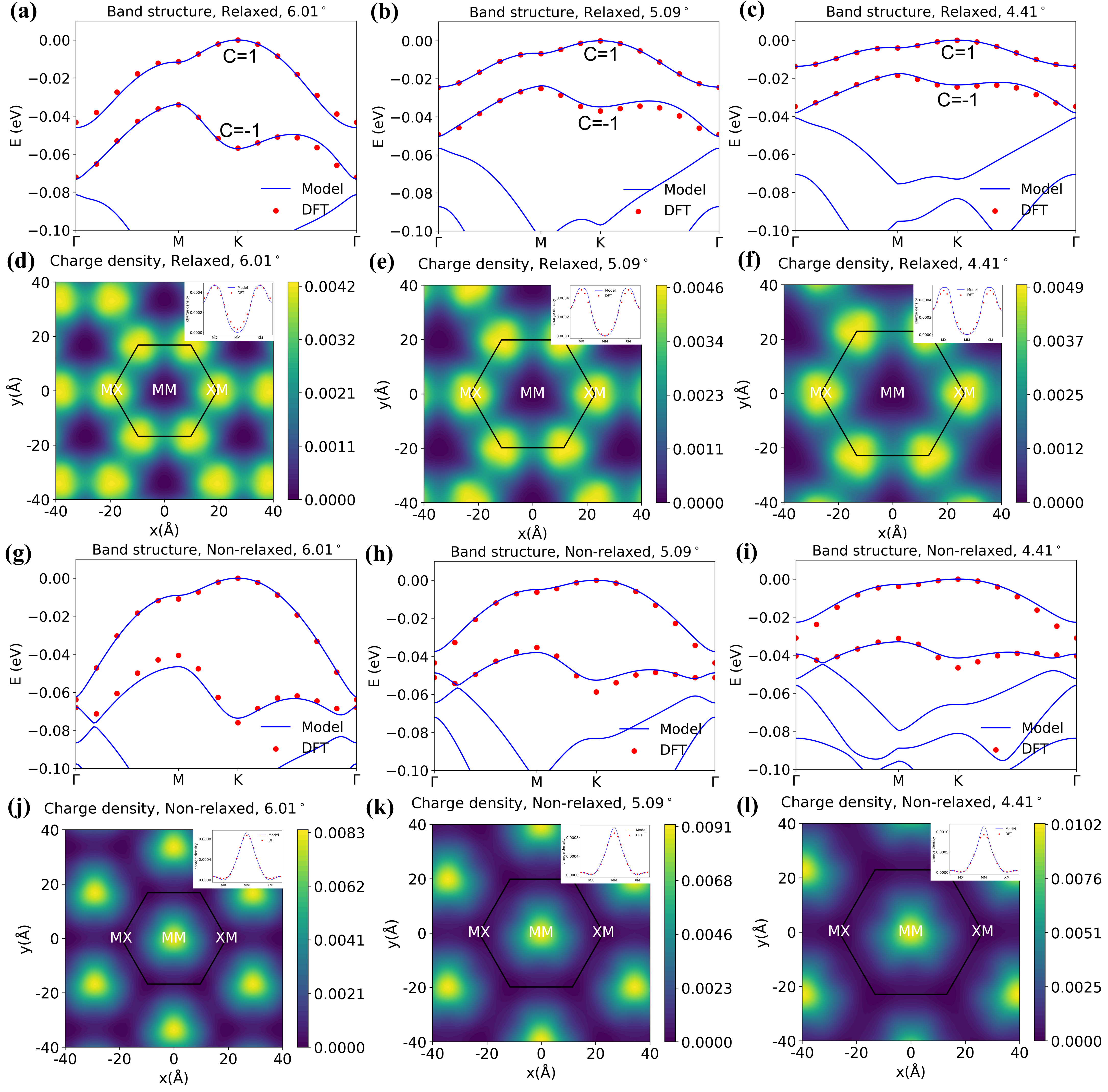}
\caption{The electronic band structures and real space distributions of  the charge density for twist MoTe$_{2}$. (a), (b),(c) The band structures for fully relaxed twisted MoTe$_{2}$ with twist angle $\theta\!=\!6.01^{\circ}$, $5.09^{\circ}$ and $4.41^{\circ}$. The red dots are band structures calculated by DFT and the blue lines are those calculated by the continuum model. The Chern number of the first moir\'e band is +1 and that of the second moir\'e band is -1. (d), (e), (f) The real space distribution of charge densities for fully relaxed twist MoTe$_{2}$ with twist angle $\theta\!=\!6.01^{\circ}$, $5.09^{\circ}$ and $4.41^{\circ}$. The inserts represent the charge densities along the high symmetry line. The red dots are band structures calculated by DFT and the blue lines are that calculated by continuum model. (g), (h), (i) The band structures for unrelaxed twisted MoTe$_{2}$. (j), (k), (l) The charge densities for unrelaxed twisted MoTe$_{2}$.}
\label{fig4}
\end{figure*}

\subsection{Numerical results}
We perform the DFT calculation using VASP package. The detailed settings are presented in Appendix. We focus on twisted bilayer MoTe$_{2}$ with twist angle $6.00^{\circ}$, $5.08^{\circ}$ and $4.40^{\circ}$. For each system, we calculate three key quantities: the band structures, the real space distribution of charge densities and Chern number, with both relaxed and unrelaxed lattice structures. For band structures, we fit the first and the second moir\'e bands calculated using the continuum model to those calculated by DFT. For charge density, we fit the real space distribution of charge densities contributed by $\Gamma$ point of the first moir\'e band. The Chern number is calculated based on the $C_3$ eigenvalues of the corresponding Bloch states at high-symmetry points, as implemented in IRVSP package \cite{irvsp}. We compare the Chern numbers of the first and second moir\'e bands calculated by continuum model and DFT, which need to be consistent with each other. 

In Fig.~\ref{fig4}, we present the band structures, the real space distribution of charge densities and Chern numbers calculated by continuum model and DFT at three different twist angles. The first and second moir\'e bands calculated by different methods exhibit nearly identical (the discrepancy is within 2.4\%) dispersion relations at three different twist angles $6.01^{\circ}$, $5.09^{\circ}$, $4.41^{\circ}$, as shown in Fig.~\ref{fig4}(a)-(c) for relaxed lattice structures, and in Fig.~\ref{fig4}(g)-(i) for unrelaxed lattice structures. The real-space distribution of charge densities calculated using continuum model also  show quantitative agreement with those from DFT, with the discrepancy less than 4.1\%, as shown in Fig.~\ref{fig4}(d)-(f) for relaxed lattice structures, and in Fig.~\ref{fig4}(j)-(l) for unrelaxed lattice structures. It indicates that our model successfully captures the norm of the low-energy wavefunctions. Besides, both continuum model and DFT calculations yield the same Chern numbers for the first and second moir\'e bands at all the three twist angles, indicating that our model accurately captures the geometric phase of the low-energy wavefunctions. The Chern number of the first moir\'e band is $+1$ and that of the second moir\'e band is $-1$ for all three twist angles with fully relaxed lattice structures.
Notably, this model performs well for both unrelaxed and relaxed lattice structures. For unrelaxed systems, the agreement reflects the model’s ability to capture intrinsic moir\'e effects; for relaxed systems, it validates the strain-induced corrections. This dual consistency provides strong evidence for the model's physical plausibility.

\begin{table}[h]
    \centering
    \begin{tabular}{l c c c}
        \hline
        Parameters & Norm & Phase ($\pi$) & Unit \\
        \hline
        $A^{l}$ & 25.10 & 0 & eV\\
        $V_{[0]}$ & 22.43 & 0.7846 & meV \\
        $V_{[0]}'$ & -0.024 & 0.7846 & meV/$\angs$ \\
         $\tilde{W}_{[0],0}$ & 37.69 & 0 & meV\\
         $\tilde{W}_{[0],1}$ & -95.90 & 0 & meV/$\angs$\\
         $\tilde{W}_{[0],2}$ & 418.96 & 0 & meV/$\angs^{2}$\\
         $V_{[1]}$ & 22.36 & 0.1759 & meV \\
         $\tilde{W}_{[1],0}$ & 29.80 & 0.1947 & meV\\
         $\tilde{W}_{[1],1}$ & -61.30 & 0.1947 & meV/$\angs$\\
         $\tilde{W}_{[1],2}$ & -300.21 & 0.1947 & meV/$\angs^{2}$\\
         $V_{[2]}$ & 40.08 & 0.5079 & meV \\
         $\tilde{W}_{[2],0}$ & 15.42 & 0.6974 & meV\\
        \hline
    \end{tabular}
    \caption{The universal parameters for twisted bilayer MoTe$_{2}$.}
    \label{table:parameter}
\end{table}

We also check the physical meaning of each parameter. In TABLE.~\ref{table:parameter}, we present the model parameters for twisted bilayer MoTe$_{2}$ constructed using our formalism. The typical value of the correction to kinetic energy induced by the lattice relaxations is about $4\,$meV. The intralayer moir\'e potential with first Fourier component $V_{[0]}$ has a magnitude of $22.43\,$meV, which does not change as $d_{0}$ varies. The typical value of strain induced correction to intralayer moir\'e potential is about $5\,$meV. The amplitude of the interlayer moir\'e potential with first Fourier component is $37.69\,$meV, which increases as $d_{0}$ decreases. The typical value of strain induced correction to interlayer moir\'e potential is about $1\,$meV.

\begin{figure*}
\includegraphics{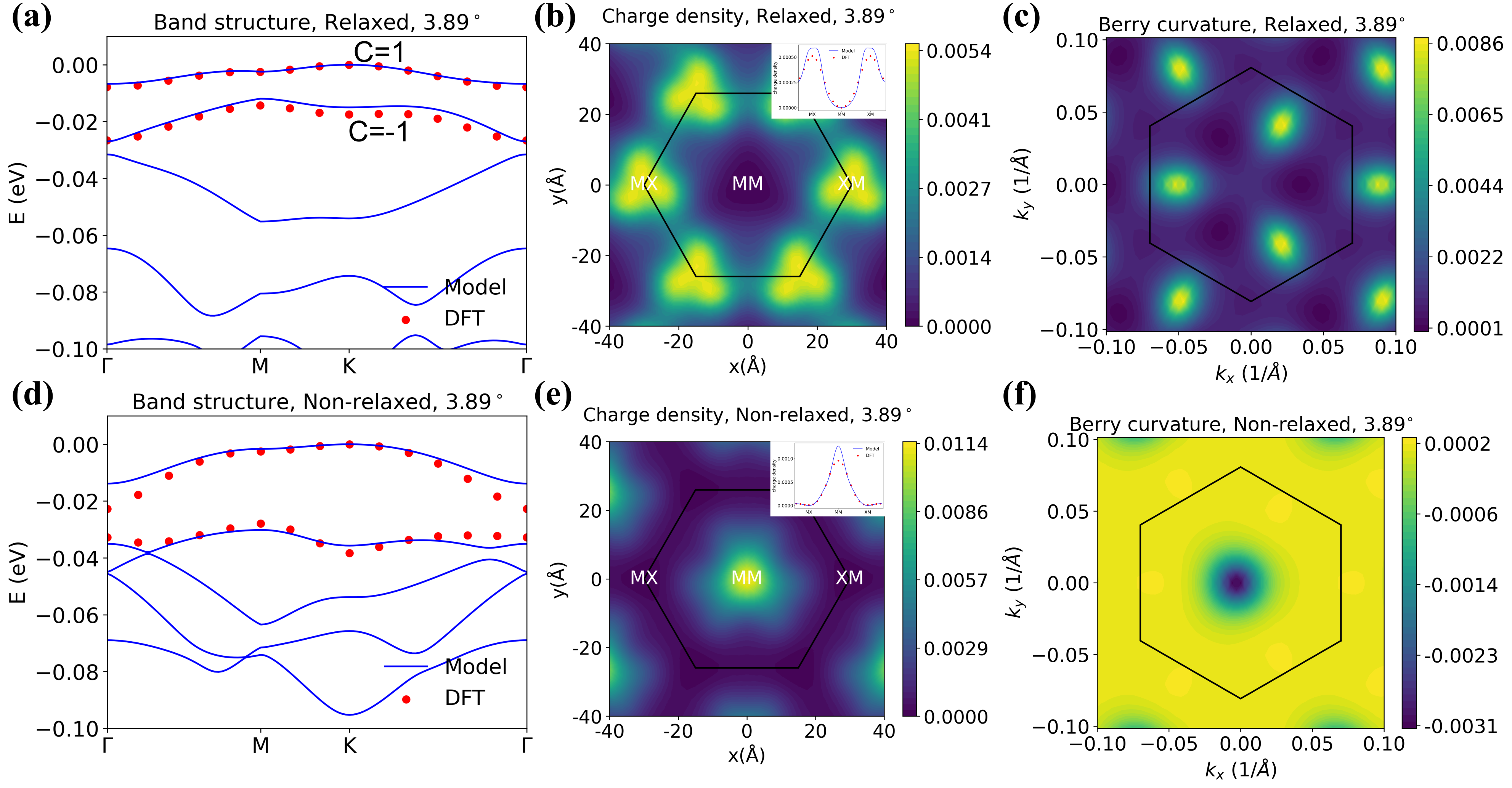}
\caption{The electronic band structures, real space distributions of the charge density and Berry curvature for twisted MoTe$_{2}$ with twist angle $\theta\!=\!3.89^{\circ}$ (not included in the fitting data). The band structures for (a) fully relaxed and (d) unrelaxed lattice structure. The red dots are band structures calculated by DFT, and the blue lines are those calculated by the continuum model. The Chern number of the first moir\'e band is +1 and that of the second moir\'e band is -1. The real space distribution of charge densities for (b) fully relaxed and (e) unrelaxed lattice structures. The inserts represent the charge densities along the high symmetry line. The red dots are charge densities calculated by DFT and the blue lines are those calculated by the continuum model. The Berry curvature distribution of the first moir\'e band with (c) fully relaxed and (f) unrelaxed lattice structures.}
\label{fig5}
\end{figure*}

A critical test of the model is its ability to predict properties of systems which are not included in the fitting dataset. In Fig.~\ref{fig5}, we present the band structures, the real space distribution of charge densities and Chern numbers calculated by continuum model and DFT at a smaller $\theta\!=\!3.89^{\circ}$, which is not included in our fitting dataset.  We find excellent agreement between the model and DFT results. As shown in Fig.~\ref{fig5}(a) and (d), the difference between band structures calculated with continuum model (blue lines) and DFT (red dots) is within 5.07\%, for both relaxed (Fig.~\ref{fig5}(a)) and unrelaxed (Fig.~\ref{fig5}(d)) lattice structures. The difference of the charge densities calculated using the two methods is within 6.9\%, as shown in Fig.~\ref{fig5}(b) and (e) for relaxed and unrelaxed structures, respectively. The Chern number of the first moir\'e band is $+1$ and that of the second moir\'e band is $-1$ with fully relaxed lattice structures, which exhibit consistency between continuum model and DFT results. In Fig.~\ref{fig5}(c) and (f), we present the Berry curvature of the first moir\'e band with fully relaxed and unrelaxed lattice structures, respectively. This strong extrapolation power arises from the model's parameterization of interlayer-distance dependence and the inclusion of various lattice relaxation effects while maintaining the coupling coefficients being twist-angle independent. This enables our model to  be generalizable to larger moir\'e unit cells. Importantly, different from the projection methods introduced in Ref.~\cite{wuquansheng_without_fit_arxiv25}, our model reduces the need for expensive large-scale DFT calculations at small twist angles, saving huge amounts of computational resources.

\section{Collusion and Discussion}
\label{sec:summary}
To conclude, in this work we have developed a universal formalism of constructing generic continuum models that are applicable for arbitrary moir\'e superlattices and are extrapolatable to any twist angles. 
This continuum model formalism combines accuracy, physical plausibility, and computational efficiency. Unlike traditional methods which usually include all the intrinsic and extrinsic effects into a series of moir\'e  potential parameters, here we clearly distinguish the intrinsic properties (the microscopic kinetic energy and atomic lattice potential) and the external inputs (the twist angle and lattice relaxation patterns). The model parameters only capture the intrinsic properties of the system, thus are independent of the twist angle and lattice relaxations. However, the lattice relaxations are still coupled to the low-energy electrons through the intrinsic coupling parameters. Such a formalism enables us to construct a universal continuum model that can accurately reproduce the low-energy properties from DFT calculations at various twist angles using a single set of parameters.
Such partition also ensures that our model not only fits numerical data but also provides insights into the origins of emergent electronic properties.
The  validity of our formalism is demonstrated using the twisted bilayer MoTe$_{2}$ system, where it reproduces DFT results for band structures, charge densities and Chern numbers with both unrelaxed and relaxed lattice structures at three different twist angles $6.01^{\circ}, 5.09^{\circ}$ and $4.41^{\circ}$. Besides, our model successfully predicts results that are quantitatively consistent with those calculated by DFT at  smaller twist angle $3.89^{\circ}$, showing strong extrapolative power of our continuum model formalism. 
This is particularly valuable for studying the small twist angle regime where supercell sizes become prohibitively large,  avoiding large-scale DFT calculations.

While validated for twist bilayer MoTe$_{2}$, the formalism is designed for broad transferability across generic moir\'e systems. We intentionally include terms that may have small effects in twist MoTe$_{2}$, but might be critical in other systems, such as nonlocal moir\'e potential.
The framework’s symmetry-based construction easily adapts to systems with different symmetries. Although demonstrated using continuum model expanded around $\mathbf{K}$/$\mathbf{K}'$ points, our theoretical formalism is completely universal and applies to continuum model construction expanded around any high-symmetry points in the atomic Brillouin zone. Generalizations to other valleys are straightforward, and the only difference comes from symmetry constraints.
 Moreover, our model offers flexibility in customization, allowing users to tailor loss functions, adjust iteration workflows or integrate with different DFT setups. Notably, the workflow supports controlled variable analysis of lattice relaxation effects, allowing users to sequentially incorporate specific relaxation components. For example, one can first fit corrections induced solely by out-of-plane distortions (fixing in-plane displacements) before gradually introducing in-plane relaxation effects, or vice versa. The accuracy and the transferability of our model would be improved with more fitting data set. 
Looking forward, our work lays a foundation for both exploring electronic correlations effects and building moir\'e superlattice databases. It provides a powerful tool for unraveling the rich physics of moir\'e systems and would accelerate the design of moir\'e-based low dimensional quantum devices.

\bibliography{tmg-phonon-tmd}

\begin{widetext}
\clearpage

\begin{center}
\textbf{\large APPENDIX}
\end{center}

\subsection{Symmetry analysis}
In each specific moir\'e superlattice system, the symmetries in the system would apply constraints on the parameters in the effective continuum model. We start with the wave function:
\begin{equation}
|\Psi_{n,\tilde{\textbf{k}}}\rangle = \sum_{\lambda} C_{\lambda, \textbf{G}+\Delta\textbf{K}_{l},n,\tilde{\textbf{k}}} |\lambda,  \tilde{\textbf{k}}+\textbf{G} +\Delta\textbf{K}_{l} \rangle
\end{equation}
Then we can write down the Hamiltonian:
\begin{align}
H =& \langle \Psi_{n,\tilde{\textbf{k}}} | \hat{H} | \Psi_{n,\tilde{\textbf{k}}}\rangle \nn
=& \sum_{\lambda, \textbf{G}} \sum_{\lambda', \textbf{G}'}  C_{\lambda', \textbf{G}'+\Delta\textbf{K}_{l'} ,n,\tilde{\textbf{k}}}^{*}C_{\lambda, \textbf{G}+\Delta\textbf{K}_{l} ,n,\tilde{\textbf{k}}} \langle \lambda', \textbf{G}' + \tilde{\textbf{k}}+\Delta\textbf{K}_{l'} | \hat{H} | \lambda, \textbf{G} + \tilde{\textbf{k}}+\Delta\textbf{K}_{l} \rangle\nn
=& \sum_{\lambda, \textbf{G}} \sum_{\lambda', \textbf{G}'} C_{\lambda', \textbf{G}'+\Delta\textbf{K}_{l'} ,n,\tilde{\textbf{k}}}^{*} \, C_{\lambda, \textbf{G}+\Delta\textbf{K}_{l} ,n,\tilde{\textbf{k}}} \, H(\tilde{\textbf{k}})_{\lambda, \textbf{G}; \lambda', \textbf{G}'},
\end{align}
where $\lambda$ represents the flavors in continuum, for example spin, valley, sublattice, etc. We analysis the influence of symmetry operator $O_g$ on the basis function form the $l$-th layer:
\begin{equation}
\langle \textbf{r} | \lambda, \textbf{G} + \tilde{\textbf{k}}-\Delta\textbf{K}_{l} \rangle = \chi_{\lambda} (\textbf{r}) \, e^{i (\textbf{G} + \tilde{\textbf{k}}+\Delta\textbf{K}_{l}) \cdot \textbf{r}},
\end{equation}
where $\Delta\textbf{K}_{l}\!=\!\textbf{K}_{\mu,l}-\textbf{K}_{\mu}$ represents the deviation from the untwist $K$ point. Under operation $O_{g}$, we transform the basis of the wave function as:
\begin{align}
\chi_{\lambda} (\textbf{r}) \, e^{i (\textbf{G} + \tilde{\textbf{k}}+\Delta\textbf{K}_{l}) \cdot \textbf{r}} &\rightarrow \sum_{\lambda'} \left[O_{g}\right]_{\lambda',\lambda} \, \chi_{\lambda'} (\textbf{r}) \,  e^{i g^{-1}(\textbf{G} + \tilde{\textbf{k}}+\Delta\textbf{K}_{l}) \cdot \textbf{r}} \nn
\langle \textbf{r} | \lambda, \textbf{G} + \tilde{\textbf{k}}+\Delta\textbf{K}_{l} \rangle &\rightarrow \sum_{\lambda'} \left[O_{g}\right]_{\lambda',\lambda} \langle \textbf{r} | \lambda', g^{-1}(\textbf{G} + \tilde{\textbf{k}}+\Delta\textbf{K}_{l}) \rangle
\end{align}
Under operation $O_{g}$, the Hamiltonian transforms as:
\begin{align}
&\langle \lambda', \textbf{G}' + \tilde{\textbf{k}}+\Delta\textbf{K}_{l'} | \hat{H} | \lambda, \textbf{G} + \tilde{\textbf{k}}+\Delta\textbf{K}_{l} \rangle =H(\tilde{\textbf{k}})_{\lambda, \textbf{G}; \lambda', \textbf{G}'}\rightarrow\nn
&\sum_{\lambda'_{1},\lambda_{1} } \left[O_{g}\right]^{*}_{\lambda'_{1},\lambda'}\langle \lambda'_{1}, g^{-1}(\textbf{G}' + \tilde{\textbf{k}}+\Delta\textbf{K}_{l'_{1}}) | \hat{H} | \lambda_{1}, g^{-1}(\textbf{G} + \tilde{\textbf{k}}+\Delta\textbf{K}_{l_{1}}) \rangle  \left[O_{g}\right]_{\lambda_{1},\lambda}= \left[ O_{g}^{\dagger} H (g^{-1}(\tilde{\textbf{k}}))_{g^{-1}(\textbf{G}),g^{-1}(\tilde{\textbf{G}'})} \, O_{g} \right]_{\lambda, \lambda'}
\end{align}
If $O_{g}$ is a symmetry of this system, we require:
\begin{equation}
\Rightarrow \left[ O_{g}^{\dagger} H (g^{-1}(\tilde{\textbf{k}}))_{g^{-1}(\textbf{G}),g^{-1}(\tilde{\textbf{G}'})} \, O_{g} \right]_{\lambda, \lambda'} = H (\tilde{\textbf{k}})_{\lambda,\textbf{G}; \lambda',\textbf{G}'}
\end{equation}

We can write down a low-energy continuum model:
\begin{equation}
H(\tilde{\textbf{k}}')_{\lambda,\textbf{G};\lambda',\textbf{G}'}=T(\tilde{\textbf{k}}+\textbf{G}+\Delta\textbf{K}_{l})_{\lambda,\lambda'}+\sum_{j}V(\tilde{\textbf{k}},\textbf{Q}_{j})_{\lambda,\lambda'},
\end{equation}
where $\textbf{Q}_{j}\!=\!\textbf{Q}^{Ref}_{j}+\tilde{\textbf{k}}+\textbf{G}'-\textbf{G}$, $\textbf{Q}^{Ref}_{j}$ is the reference k point.

For the kinetic energy, we can expand with respect to momentum:
\begin{equation}
T(\tilde{\textbf{k}}+\textbf{G}+\Delta\textbf{K}_{l})_{\lambda,\lambda'}=\sum^{\infty}_{\alpha=0}\sum^{\alpha}_{\beta=0}\bar{T}^{\alpha,\beta}_{\lambda,\lambda'}(\tilde{\textbf{k}}+\textbf{G}+\Delta\textbf{K}_{l})_{x}^{\beta}(\tilde{\textbf{k}}+\textbf{G}+\Delta\textbf{K}_{l})_{y}^{\alpha-\beta}.
\end{equation}
Under symmetry operation $O_{g}$, the kinetic energy transforms as:
\begin{align}
&T(\tilde{\textbf{k}}+\textbf{G}+\Delta\textbf{K}_{l})_{\lambda,\lambda'}\rightarrow\nn
 &\left[\left[O_{g}\right]^{\dagger}T(g^{-1}(\tilde{\textbf{k}}+\textbf{G}+\Delta\textbf{K}_{l}))O_{g} \right]_{\lambda,\lambda'}\nn
 &=\sum^{\infty}_{\alpha=0}\sum^{\alpha}_{\beta=0}\sum_{\lambda_{1},\lambda'_{1}}\left[O_{g}\right]^{*}_{\lambda'_{1},\lambda'}\bar{T}^{\alpha,\beta}_{\lambda_{1},\lambda'_{1}}\left[O_{g}\right]_{\lambda_{1},\lambda}g^{-1}(\tilde{\textbf{k}}+\textbf{G}+\Delta\textbf{K}_{l})_{x}^{\beta}g^{-1}(\tilde{\textbf{k}}+\textbf{G}+\Delta\textbf{K}_{l})_{y}^{\alpha-\beta}.
\end{align}
If $O_{g}$ is a symmetry, we require:
\begin{equation}
\sum^{\infty}_{\alpha=0}\sum^{\alpha}_{\beta=0}\bar{T}^{\alpha,\beta}_{\lambda,\lambda'}(\tilde{\textbf{k}}+\textbf{G}+\Delta\textbf{K}_{l})_{x}^{\beta}(\tilde{\textbf{k}}+\textbf{G}+\Delta\textbf{K}_{l})_{y}^{\alpha-\beta}=\sum^{\infty}_{\alpha=0}\sum^{\alpha}_{\beta=0}\sum_{\lambda_{1},\lambda'_{1}}\left[O_{g}\right]^{*}_{\lambda'_{1},\lambda'}\bar{T}^{\alpha,\beta}_{\lambda_{1},\lambda'_{1}}\left[O_{g}\right]_{\lambda_{1},\lambda}g^{-1}(\tilde{\textbf{k}}+\textbf{G}+\Delta\textbf{K}_{l})_{x}^{\beta}g^{-1}(\tilde{\textbf{k}}+\textbf{G}+\Delta\textbf{K}_{l})_{y}^{\alpha-\beta}
\end{equation}

For the potential term, we expand with respect to the reference k point:
\begin{equation}
\sum_{i,j}V(\tilde{\textbf{k}},\textbf{Q}_{[i],j})_{\lambda,\lambda'}=\sum_{j}\sum^{\infty}_{\alpha=0}\sum^{\alpha}_{\beta=0}\bar{V}^{\alpha,\beta}_{\lambda,\lambda'}(\textbf{Q}_{[i],j})\cdot(\tilde{\textbf{k}}+\textbf{G})_{x}^{\beta}(\tilde{\textbf{k}}+\textbf{G})_{y}^{\alpha-\beta},
\end{equation}
where $\bar{V}^{\alpha,\beta}_{\lambda,\lambda'}(\textbf{Q}_{[i],j})$ is independent of momentum. We can further expand $\bar{V}$ with respect to the strain fields:
\begin{equation}
\bar{V}^{\alpha,\beta}_{\lambda,\lambda'}(\textbf{Q}_{[i],j})\approx \bar{V}^{\alpha,\beta,0}_{\lambda,\lambda'}(\textbf{Q}_{[i],j})+\sum_{s}\sum_{\textbf{G}_{s}}\left[\frac{\partial}{\partial S^{s}_{\textbf{G}_{s}}}\bar{V}^{\alpha,\beta}_{\lambda,\lambda'}(\textbf{Q}_{[i],j})\right]S_{\textbf{G}_{s}}^{s}+\sum_{s_{1},s_{2}}\sum_{\textbf{G}_{s_{1}},\textbf{G}_{s_{2}}}\left[\frac{\partial^{2}}{\partial S^{s_{1}}_{\textbf{G}_{s_{1}}}\partial S^{s_{2}}_{\textbf{G}_{s_{2}}}}\bar{V}^{\alpha,\beta}_{\lambda,\lambda'}(\textbf{Q}_{[i],j})\right]S_{\textbf{G}_{s_{1}}}^{s_{1}}S_{\textbf{G}_{s_{2}}}^{s_{2}}+\dots,
\end{equation}
where $S=\{\textbf{u}^{\pm}, h^{\pm}\}$. We can evaluate the constraint of symmetry on the potential energy. For the zero-th order:
\begin{align}
&\sum_{j}\sum_{\alpha,\beta}\bar{V}^{\alpha,\beta,0}_{\lambda,\lambda'}(\textbf{Q}_{[i],j})\cdot(\tilde{\textbf{k}}+\textbf{G})_{x}^{\beta}(\tilde{\textbf{k}}+\textbf{G})_{y}^{\alpha-\beta}\rightarrow\nn
&\sum_{j}\sum_{\alpha,\beta}\sum_{\lambda_{1},\lambda'_{1}}\left[O_{g}\right]^{\dagger}_{\lambda'_{1},\lambda'}\bar{V}^{\alpha,\beta,0}_{\lambda_{1},\lambda'_{1}}(g^{-1}(\textbf{Q}_{[i],j}))\left[O_{g}\right]_{\lambda_{1},\lambda}\cdot g^{-1}(\tilde{\textbf{k}}+\textbf{G})_{x}^{\beta}g^{-1}(\tilde{\textbf{k}}+\textbf{G})_{y}^{\alpha-\beta}.
\end{align}
For the first order term, the strain fields are coupled to the Hamiltonian through a deformation potential. The corresponding symmetry operator on the strain field is noted as $\tilde{g}$. Under symmetry operation $O_{g}$, we have:
\begin{align}
&\sum_{j}\sum_{\alpha,\beta}\sum_{s}\sum_{\textbf{G}_{s}}\left[\frac{\partial}{\partial S^{s}_{\textbf{G}_{s}}}\bar{V}^{\alpha,\beta}_{\lambda,\lambda'}(\textbf{Q}_{[i],j})\right]\cdot S^{s}_{\textbf{G}_{s}}(\tilde{\textbf{k}}+\textbf{G})_{x}^{\beta}(\tilde{\textbf{k}}+\textbf{G})_{y}^{\alpha-\beta}\rightarrow\nn
&\sum_{j}\sum_{\alpha,\beta}\sum_{\lambda_{1},\lambda'_{1}}\sum_{s'}\left[\frac{\partial}{\partial S^{s}_{\textbf{G}_{s}}}\left[O_{g}\right]^{\dagger}_{\lambda'_{1},\lambda'}\bar{V}^{\alpha,\beta}_{\lambda_{1},\lambda'_{1}}(g^{-1}(\textbf{Q}_{[i],j}))\left[O_{g}\right]_{\lambda_{1},\lambda}\right]\cdot \tilde{g}_{s,s'} S^{s'}_{\textbf{G}_{s}} g^{-1}(\tilde{\textbf{k}}+\textbf{G})_{x}^{\beta}g^{-1}(\tilde{\textbf{k}}+\textbf{G})_{y}^{\alpha-\beta}.
\end{align}
For the second order term:
\begin{align}
&\sum_{j}\sum_{\alpha,\beta}\sum_{s_{1},s_{2}}\sum_{\textbf{G}_{s_{1}},\textbf{G}_{s_{2}}}\left[\frac{\partial^{2}}{\partial S^{s_{1}}_{\textbf{G}_{s_{1}}}\partial S^{s_{2}}_{\textbf{G}_{s_{2}}}}\bar{V}^{\alpha,\beta}_{\lambda,\lambda'}(\textbf{Q}_{[i],j})\right]\cdot S^{s_{1}}_{\textbf{G}_{s_{1}}}S^{s_{2}}_{\textbf{G}_{s_{2}}}(\tilde{\textbf{k}}+\textbf{G})_{x}^{\beta}(\tilde{\textbf{k}}+\textbf{G})_{y}^{\alpha-\beta}\rightarrow\nn
&\sum_{j}\sum_{\alpha,\beta}\sum_{\lambda_{1},\lambda'_{1}}\sum_{s'}\left[\frac{\partial^{2}}{\partial S^{s_{1}}_{\textbf{G}_{s_{1}}}\partial S^{s_{2}}_{\textbf{G}_{s_{2}}}}\left[O_{g}\right]^{\dagger}_{\lambda'_{1},\lambda'}\bar{V}^{\alpha,\beta}_{\lambda_{1},\lambda'_{1}}(g^{-1}(\textbf{Q}_{[i],j}))\left[O_{g}\right]_{\lambda_{1},\lambda}\right]\cdot \nn
&\ \ \ \ \ \ \ \ \ \ \tilde{g}_{s_{1},s'_{1}} S^{s'_{1}}_{\textbf{G}_{s_{1}}}\tilde{g}_{s_{2},s'_{2}}S^{s'_{2}}_{\textbf{G}_{s_{2}}} g^{-1}(\tilde{\textbf{k}}+\textbf{G})_{x}^{\beta}g^{-1}(\tilde{\textbf{k}}+\textbf{G})_{y}^{\alpha-\beta}.
\end{align}

\subsection{Correction to kinetic energy induced by strain for twist MoTe$_{2}$}
We analyze the strain-induced corrections to the kinetic energy in twisted MoTe$_{2}$ systems. For twist MoTe$_{2}$ systems, we model the electronic structure using a single Wannier function per atomic unit cell in each monolayer MoTe$_{2}$. The lattice vector is defined as: $\textbf{a}_{1}\!=\!a(1,0,0)$, and $\textbf{a}_{2}\!=\!a(1/2,\sqrt{3}/2,0)$. Due to the absence of sublattice degrees of freedom in our model, we focus on the hopping events from one atomic unit cell to six nearest neighbor atomic unit cell, denoted as $\textbf{r}_{1}\!=\!a(1,0,0)$, $\textbf{r}_{2}\!=\!a(1/2,\sqrt{3}/2,0)$, $\textbf{r}_{3}\!=\!a(-1/2,\sqrt{3}/2,0)$, $\textbf{r}_{4}\!=\!a(-1,0,0)$, $\textbf{r}_{5}\!=\!a(-1/2,-\sqrt{3}/2,0)$, and $\textbf{r}_{6}\!=\!a(1/2,-\sqrt{3}/2,0)$. We introduce the three dimensional lattice strain $\hat{u}$:
\begin{align}
\begin{split}
\hat{u}=\left[\begin{array}{ccc}
				u_{xx} & u_{xy} & u_{hx} \\
				u_{xy} & u_{yy} & u_{hy} \\
				u_{hx} & u_{hy} & u_{hh} 
\end{array}\right],
\end{split}
\end{align}
where $u_{\alpha\beta}\!=\!(\partial u_{\alpha}/\partial r_{\beta}+\partial u_{\beta}/\partial r_{\alpha})/2$, $\alpha,\beta\!=\!x, y$ and $u_{h\alpha}\!=\!\partial h/\partial r_{\alpha}$, $\alpha\!=\!x, y$. Under the lattice distortion operation, the hopping vectors are modified from $\textbf{r}_{i}$ to $\textbf{r}'_{i}\!=\!(1+\hat{u})\textbf{r}_{i}$. This change in hopping geometry alters the hopping amplitude, which depends on the distance and orientation of the hopping vector. The strain-induced shift in hopping amplitude, can be expressed as:
 \allowdisplaybreaks
\begin{align}
\delta t(\textbf{r}_{i})=t(\textbf{r}'_{i})-t(\textbf{r}_{i})&\approx \left.\frac{\partial t(\textbf{r})}{\partial r_{x}}\right|_{\textbf{r}_{i}}\delta r_{i, x} + \left.\frac{\partial t(\textbf{r})}{\partial r_{y}}\right|_{\textbf{r}_{i}}\delta r_{i, y} + \left.\frac{\partial t(\textbf{r})}{\partial r_{z}}\right|_{\textbf{r}_{i}}\delta r_{i, z} \nn
&=\left.\frac{\partial t(\textbf{r})}{\partial r_{x}}\right|_{\textbf{r}_{i}}(u_{xx}r_{i, x}+u_{xy}r_{i, y}+u_{hx}r_{i, z})\nn
&+\left.\frac{\partial t(\textbf{r})}{\partial r_{y}}\right|_{\textbf{r}_{i}}(u_{xy}r_{i, x}+u_{yy}r_{i, y}+u_{hy}r_{i, z})\nn
&+\left.\frac{\partial t(\textbf{r})}{\partial r_{z}}\right|_{\textbf{r}_{i}}(u_{hx}r_{i, x}+u_{hy}r_{i, y}+u_{hh}r_{i, z}).
\end{align}
To connect this to the kinetic energy, we start with the tight-binding Hamiltonian for a single layer, $g(\textbf{k})\!=\!e^{i\textbf{k}\cdot\textbf{r}}$, where the sum runs over nearest neighbors. Focusing on low-energy states near the valence band maximum (VBM) at $\textbf{k}\approx\textbf{K}\!=\!(-4\pi/3a,0,0)$, we expand $g(\textbf{k})$ around $\textbf{k}\!=\!\textbf{K}+\tilde{\textbf{k}}$, where $\tilde{\textbf{k}}$ is a small wavevector deviation from the VBM. This expansion yields:
 \allowdisplaybreaks
\begin{align}
g(\tilde{\textbf{k}}+\textbf{K})&=\sum_{j}(t_{0}(\textbf{r}_{j})+\delta t(\textbf{r}_{j}))e^{i(\textbf{K}+\tilde{\textbf{k}})\cdot\textbf{r}_{j}}\nn
&\approx\frac{\hbar^{2}\tilde{\textbf{k}}^{2}}{2m^{*}}+\sum_{j}\delta t(\textbf{r}_{j})e^{i\textbf{K}\cdot\textbf{r}_{j}}e^{i\tilde{\textbf{k}}\cdot\textbf{r}_{j}}\nn
&\approx\frac{\hbar^{2}\tilde{\textbf{k}}^{2}}{2m^{*}}+\sum_{j}\delta t(\textbf{r}_{j})e^{i\textbf{K}\cdot\textbf{r}_{j}}(1+i\textbf{k}\cdot\textbf{r}_{j}+(i\textbf{k}\cdot\textbf{r}_{j})^{2})
\end{align}
where the first term is the bare kinetic energy. 

We evaluate the $\textbf{k}$-independent term.
 \allowdisplaybreaks
\begin{align}
&\sum_{j}\delta t(\textbf{r}_{j})e^{i\textbf{K}\cdot\textbf{r}_{j}}\nn
=&\left(-\frac{1}{2}+\frac{\sqrt{3}}{2}i\right)\left[ \left.\frac{\partial t(\textbf{r})}{\partial r_{x}}\right|_{\textbf{r}_{i}}(a\,u^{(l)}_{xx}\pm\,z\,u^{(l)}_{hx})+ \left.\frac{\partial t(\textbf{r})}{\partial r_{y}}\right|_{\textbf{r}_{1}}(a\,u^{(l)}_{xy}+^{(l)}z\,u^{(l)}_{hy})+ \left.\frac{\partial t(\textbf{r})}{\partial r_{z}}\right|_{\textbf{r}_{1}}\,u^{(l)}_{hx}\right]\nn
&+\dots\nn
=&A^{(l)}\,u^{(l)}_{xx}+2B^{(l)}\,u^{(l)}_{xy}+C^{(l)}\,u^{(l)}_{yy}+D^{(l)}\,u^{(l)}_{hx}+E^{(l)}\,u^{(l)}_{hy}\nn
=&\sum_{\textbf{G}}[i\,A^{(l)}\,G_{x}u^{(l)}_{\textbf{G},x}+i\,B^{(l)}\,G_{x}u^{(l)}_{\textbf{G},y}+i\,B^{(l)}\,G_{y}u^{(l)}_{\textbf{G},x}+i\,C^{(l)}\,G_{y}u^{(l)}_{\textbf{G},y}+i\,D^{(l)}\,G_{x}h^{(l)}_{\textbf{G}}+i\,E^{(l)}\,G_{y}h^{(l)}_{\textbf{G}}]e^{i\textbf{G}\cdot\textbf{r}}\nn
=&\sum_{\textbf{G}}i\left[\begin{matrix}
G_{x} & G_{y}
\end{matrix}
\right]\left[\begin{matrix}
A^{(l)} & B^{(l)}\\
B^{(l)} & C^{(l)}
\end{matrix}
\right]\left[\begin{matrix}
u^{(l)}_{\textbf{G},x}\\
u^{(l)}_{\textbf{G},y} 
\end{matrix}
\right]e^{i\textbf{G}\cdot\textbf{r}}+i(D^{(l)}G_{x}+E^{(l)}G_{y})h^{(l)}_{\textbf{G}}e^{i\textbf{G}\cdot\textbf{r}}\nn
=&\sum_{\textbf{G}}-i\left[\begin{matrix}
G_{x} & G_{y}
\end{matrix}
\right]\left[\begin{matrix}
A^{(l),*} & B^{(l),*}\\
B^{(l),*} & C^{(l),*}
\end{matrix}
\right]\left[\begin{matrix}
u^{(l),*}_{\textbf{G},x}\\
u^{(l),*}_{\textbf{G},y} 
\end{matrix}
\right]e^{-i\textbf{G}\cdot\textbf{r}}-i(D^{(l),*}G_{x}+E^{(l),*}G_{y})h^{(l),*}_{\textbf{G}}e^{-i\textbf{G}\cdot\textbf{r}}\nn
=&\sum_{\textbf{G}}i\left[\begin{matrix}
-G_{x} & -G_{y}
\end{matrix}
\right]\left[\begin{matrix}
A^{(l),*} & B^{(l),*}\\
B^{(l),*} & C^{(l),*}
\end{matrix}
\right]\left[\begin{matrix}
u^{(l)}_{-\textbf{G},x}\\
u^{(l)}_{-\textbf{G},y} 
\end{matrix}
\right]e^{i(-\textbf{G})\cdot\textbf{r}}+i(D^{(l),*}(-G_{x})+E^{(l)}(-G_{y}))h^{(l),*}_{\textbf{G}}e^{i(-\textbf{G})\cdot\textbf{r}},
\end{align}
as a result, ${A^{(l)}, B^{(l)}, C^{(l)}, D^{(l)}, E^{(l)}}\in \mathbb{R}$. Then we apply $C_{3z}$ rotational symmetry:
\begin{align}
&\sum_{\textbf{G}}i\left[\begin{matrix}
G_{x} & G_{y}
\end{matrix}
\right]C_{3}^{T}\left[\begin{matrix}
A^{(l)} & B^{(l)}\\
B^{(l)} & C^{(l)}
\end{matrix}
\right]C_{3}\left[\begin{matrix}
u^{(l)}_{\textbf{G},x}\\
u^{(l)}_{\textbf{G},y} 
\end{matrix}
\right]e^{i(C_{3}\textbf{G})\cdot\textbf{r}}+i(D^{(l)}(C_{3}\textbf{G})_{x}+E^{(l)}(C_{3}\textbf{G})_{y})h^{(l)}_{C_{3}\textbf{G}}e^{i(C_{3}\textbf{G})\cdot\textbf{r}}\nn
=&\sum_{\textbf{G}}i\left[\begin{matrix}
G_{x} & G_{y}
\end{matrix}
\right]\left[\begin{matrix}
A^{(l)}/4+\sqrt{3}B^{(l)}/2+3C/4 & -\sqrt{3}A^{(l)}/4-B^{(l)}/2+\sqrt{3}C^{(l)}/4\\
-\sqrt{3}A^{(l)}/4-B^{(l)}/2+\sqrt{3}C^{(l)}/4 & 3A^{(l)}/4-\sqrt{3}B^{(l)}/2+C^{(l)}/4
\end{matrix}
\right]\left[\begin{matrix}
u^{(l)}_{\textbf{G},x}\\
u^{(l)}_{\textbf{G},y} 
\end{matrix}
\right]e^{i(C_{3}\textbf{G})\cdot\textbf{r}}\nn
&+i(D^{(l)}(C_{3}\textbf{G})_{x}+E^{(l)}(C_{3}\textbf{G})_{y})h^{(l)}_{C_{3}\textbf{G}}e^{i(C_{3}\textbf{G})\cdot\textbf{r}}.
\end{align}
$C_{3}$ symmetry enforces $A^{(l)}\!=\!C^{(l)}$ and $B^{(l)}\!=\!0$. Then we consider $C_{2y}T$ symmetry, it would flip two layers in this system.
\begin{align}
&\sum_{\textbf{G}}-i(C_{2y}\textbf{G})\cdot\textbf{u}^{(-l),*}_{C_{2y}\textbf{G}}A^{(-l)}e^{-i(C_{2y}\textbf{G})\cdot\textbf{r}}\nn
=&\sum_{\textbf{G}}i(-C_{2y}\textbf{G})\cdot\textbf{u}^{(-l)}_{-C_{2y}\textbf{G}}A^{(-l)}e^{i(-C_{2y}\textbf{G})\cdot\textbf{r}}.
\end{align}
$C_{2y}T$ symmetry enforces $A^{(l)}\!=\!A^{(-l)}$. For system with $C_{3z}$ and $C_{2y}T$ symmetry, the correction to kinetic energy induced by strain can be expressed as:
\begin{align}
&\sum_{j}\delta t(\textbf{r}_{j})=\sum_{\textbf{G}}i\left[\begin{matrix}
G_{x} & G_{y}
\end{matrix}
\right]\left[\begin{matrix}
u^{(l)}_{\textbf{G},x}\\
u^{(l)}_{\textbf{G},y} 
\end{matrix}
\right]A^{(l)}e^{i\textbf{G}\cdot\textbf{r}},\nn
&\bra{\tilde{\textbf{k}}+\textbf{G}',l}V_{s}\ket{\tilde{\textbf{k}}+\textbf{G},l}=\sum_{\textbf{G}_{s}}i\left[\begin{matrix}
G_{s, x} & G_{s, y}
\end{matrix}
\right]\left[\begin{matrix}
u^{(l)}_{\textbf{G}_{s},x}\\
u^{(l)}_{\textbf{G}_{s},y} 
\end{matrix}
\right]A^{(l)}\delta_{\textbf{G}',\textbf{G}+\textbf{G}_{s}}
\end{align}

\subsection{Non-local moir\'e potential}
In this section, we analyze the nonlocal moir\'e potential in the generic continuum model. In previous studies, the nonlocal moiré potential in twisted bilayer graphene was treated as a $\textbf{k}$-dependent correction to the bare moir\'e potential \cite{Song-tbg2-arxiv20}. Here we consider the most general form of the nonlocal moir\'e potential. 
We start with the Fourier transformation of the moir\'e potential:
\begin{align}
V(\textbf{r},\textbf{r}')=\sum_{\tilde{\textbf{k}},\textbf{G}}\sum_{\tilde{\textbf{k}}',\textbf{G}'}e^{-i(\tilde{\textbf{k}}+\textbf{G})\cdot\textbf{r}}V_{\textbf{G},\textbf{G}'}(\tilde{\textbf{k}},\tilde{\textbf{k}}')e^{i(\tilde{\textbf{k}}'+\textbf{G}')\cdot\textbf{r}'}.
\end{align}
The inverse Fourier transform is 
\begin{align}
V_{\textbf{G},\textbf{G}'}(\tilde{\textbf{k}},\tilde{\textbf{k}}')=\frac{1}{S^{2}}\int_{\Omega_{M}}\mathrm{d}^{2}\tilde{r}\int_{\Omega_{M}}\mathrm{d}^{2}\tilde{r}'\,\sum_{\textbf{R}}\sum_{\textbf{R}'}e^{i(\tilde{\textbf{k}}+\textbf{G})\cdot(\tilde{\textbf{r}}+\textbf{R})}V(\tilde{\textbf{r}}+\textbf{R},\tilde{\textbf{r}}'+\textbf{R}')e^{-i(\tilde{\textbf{k}}'+\textbf{G}')\cdot(\tilde{\textbf{r}}'+\textbf{R}')}
\end{align}
where $S=N\Omega_{M}$ is the total area of the system, $\Omega_M$ is the area of the moir\'e unit cell, and $N$ is the number of moir\'e cells in the entire system.

We have the translational symmetry $V(\tilde{\textbf{r}}+\textbf{R},\tilde{\textbf{r}}'+\textbf{R}')\!=\!V(\tilde{\textbf{r}}+\textbf{R}+\textbf{R}_{0},\tilde{\textbf{r}}'+\textbf{R}'+\textbf{R}_{0})$, so that $V(\tilde{\textbf{r}}+\textbf{R},\tilde{\textbf{r}}'+\textbf{R}')\!=\!V(\tilde{\textbf{r}}+\textbf{R}-\textbf{R}',\tilde{\textbf{r}}')$. We have:
\begin{align}
&V_{\textbf{G},\textbf{G}'}(\tilde{\textbf{k}},\tilde{\textbf{k}}')\nn
=&\frac{1}{N^{2}\Omega^{2}_{M}}\sum_{\textbf{R},\textbf{R}'}\int_{\Omega_{M}}\mathrm{d}^{2}\tilde{r}\int_{\Omega_{M}}\mathrm{d}^{2}\tilde{r}'\,e^{i(\tilde{\textbf{k}}+\textbf{G})\cdot(\tilde{\textbf{r}}+\textbf{R})}V(\tilde{\textbf{r}}+\textbf{R}-\textbf{R}',\tilde{\textbf{r}})e^{-i(\tilde{\textbf{k}}'+\textbf{G}')\cdot(\tilde{\textbf{r}}'+\textbf{R}')}\nn
=&\frac{1}{N^{2}\Omega^{2}_{M}}\sum_{\Delta\textbf{R},\textbf{R}'}\int_{\Omega_{M}}\mathrm{d}^{2}\tilde{r}\int_{\Omega_{M}}\mathrm{d}^{2}\tilde{r}'\,e^{i\tilde{\textbf{k}}\cdot\Delta\textbf{R}}e^{i(\tilde{\textbf{k}}-\tilde{\textbf{k}}')\cdot\textbf{R}'}e^{i((\tilde{\textbf{k}}+\textbf{G})\cdot\tilde{\textbf{r}}}V(\tilde{\textbf{r}}+\Delta\textbf{R},\tilde{\textbf{r}}')e^{-i(\tilde{\textbf{k}}'+\textbf{G}')\cdot\tilde{\textbf{r}}'}\nn
=&\delta_{\tilde{\textbf{k}},\tilde{\textbf{k}}'}\frac{1}{N}\sum_{\Delta\textbf{R}}\frac{1}{\Omega^{2}_{M}}\int_{\Omega_{M}}\mathrm{d}^{2}\tilde{r}\int_{\Omega_{M}}\mathrm{d}^{2}\tilde{r}'\,e^{i(\tilde{\textbf{k}}+\textbf{G})\cdot(\tilde{\textbf{r}}+\Delta\textbf{R})}V(\tilde{\textbf{r}}+\Delta\textbf{R},\tilde{\textbf{r}}')e^{-i(\tilde{\textbf{k}}+\textbf{G}')\cdot\tilde{\textbf{r}}'}\nn
=&\delta_{\tilde{\textbf{k}},\tilde{\textbf{k}}'}\frac{1}{N}\sum_{\Delta\textbf{R}}\frac{1}{\Omega^{2}_{M}}\int_{\Omega_{M}}\mathrm{d}^{2}\tilde{r}\int_{\Omega_{M}}\mathrm{d}^{2}\tilde{r}'\,e^{i(\tilde{\textbf{k}}+\textbf{G})\cdot(\tilde{\textbf{r}}+\Delta\textbf{R}-\tilde{\textbf{r}}')}V(\tilde{\textbf{r}}+\Delta\textbf{R},\tilde{\textbf{r}}')e^{-i(\textbf{G}'-\mathbf{G})\cdot\tilde{\textbf{r}}'}\;\nn
=&\delta_{\tilde{\mathbf{k}},\tilde{\mathbf{k}}'}\,V_{\mathbf{G}'-\mathbf{G}}(\tilde{\mathbf{k}}+\mathbf{G})
\end{align}

With fixed $\mathbf{G'}-\mathbf{G}$, $V_{\mathbf{G}'-\mathbf{G}}(\tilde{\mathbf{k}}+\mathbf{G})$ can be expanded in terms of $\mathbf{k}=\tilde{\mathbf{k}}+\mathbf{G}$
\begin{align}
&V_{\mathbf{G}'-\mathbf{G}}(\tilde{\mathbf{k}}+\mathbf{G})\nn
=&\sum_{n,m}\,V_{\textbf{G}-\textbf{G}'}^{n,m}\,\mathcal{T}_{n}(\left|\textbf{k}\right|L_{s}\eta)e^{im\phi}.
\end{align}
where $\mathbf{k}=\vert\mathbf{k}\vert(\cos{\phi},\sin{\phi})$ is expressed in polar coordinate.
Here $\mathcal{T}_{n}$ is the $n$-th order of Chebyshev polynomial. The factor $\eta$ a free parameter to be determined. In principle, each expansion coefficient $V_{\textbf{G}-\textbf{G}'}^{n,m}$ is complex valued. To reduce the number of parameters (and computational cost in fitting these parameters), in practical calculations we make the following approximation:
\begin{align}
&V_{\mathbf{G}'-\mathbf{G}}(\tilde{\mathbf{k}}+\mathbf{G})\nn
\approx&\sum_{n,m}\,\alpha_{\textbf{G}-\textbf{G}'}^{n,m}\,\mathcal{T}_{n}(\left|\textbf{k}\right|L_{s}\eta)\cos{(m\phi+\phi_m)} e^{i\psi}\;.
\end{align}
Here $\alpha_{\textbf{G}-\textbf{G}'}^{n,m}$ is real valued, having a constant phase angle $e^{i\psi}$. $\phi_m$ is additional set of parameters to be determined in the fitting process. In our calculations, we only include the nonlocal moir\'e potential terms to the first Fourier component, with $\mathbf{G}'-\mathbf{G}=\mathbf{Q}_j$, $\mathbf{Q}_j$ ($j=1, 2, 3$) denotes the three primitive reciprocal vectors of the moir\'e Brillouin zone. 


\subsection{Details in machine learning force field setups}

In this section, we present the details of machine learning force field (MLFF). We are using the DeePMD-kit code to train the neural network \cite{dpmd, dpmd2}. 4000 untwist $3\times3$ bilayer MoTe$_{2}$ are constructed with each perturbed with random change in lattice constant, interlayer and intralayer shift, and atomic displacement to capture the various local configrations in the moir\'e superlattice. The training dataset consists of the total energy, atomic force and virial tensor of the 4000 frames. We adopt a smooth edition of the two-atom embedding descriptor 'se\_e2\_a', which is constructed from both angular and radial informations of atomic configurations. The cut-off radius for neighbor searching is set as 10 Å and the smoothing starts from 0.5 Å, which gives a maximum number of 64 Mo and 136 Te atoms. We construct a embedding network that possesses three hidden layers with (30,60,120) neurons in each of them, respectively. The fitting network employs a uniform architecture, with hidden layers configured as (240,240,240). To measure the quality of the neural network, we have fully taken into account the information in the training data to construct the following loss function:
\begin{align}
L\left(p_\varepsilon,p_f,p_\xi\right)=\frac{p_\varepsilon}{N}\Delta\varepsilon^2+\frac{p_f}{3N}\sum_i\left | \Delta\mathbf{F}_i \right |^2+\frac{p_\xi}{9N}\left \| \Delta\xi \right \|^2
\end{align}
where $\Delta\varepsilon$, $\Delta\mathbf{F}_i$, and $\Delta\xi$ denote root mean square error (RMSE) in energy, force, and virial, respectively. Using the trained MLFFs, we perform large-scale lattice relaxation simulations using Large-scale Atomic/Molecular Massively Parallel Simulator (LAMMPS) package \cite{lammps}. 

In Fig. SI~\ref{mlff}, we present the validation of the neural network potential. We randomly select 100 frames from the training dataset as the testing dataset to check the convergence of energy, force, and stress. The energy and force testing error between the deep potential method and DFT dataset are less than 2 meV/atom and 100 meV/Å, respectively, which indicates a good accuracy of the MLFF.
We also evaluate the performance of the trained MLFF by relaxing twist MoTe$_{2}$ with $\theta\!=\!6.01^{\circ}$, and compare the structure with that relaxed using DFT. The maximum differences in atomic positions are found to be 0.0369$\,\angs$, which is much smaller than the amplitudes of relaxation fields.

\begin{FIGSI}
\includegraphics[width=8cm]{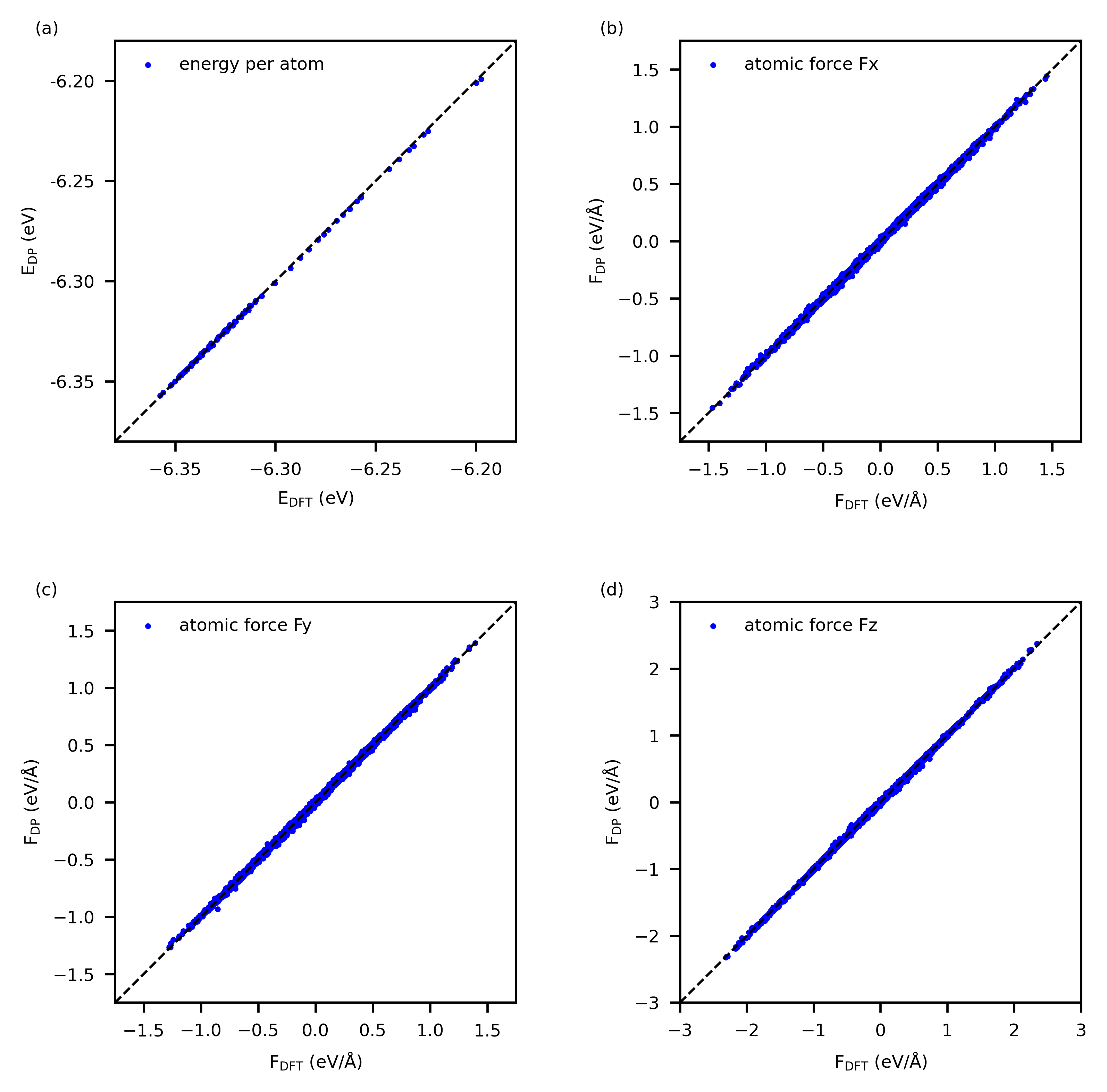}
\caption{The validation of the neural network potential for total energy and force along three directions. }
\label{mlff}
\end{FIGSI}

\subsection{Details in first principle calculation}

In this section, we present the details of the first principles calculation. The training dataset is obtained by the self-consistent calcualtion using the tool of Vienna ab initio simulation package (VASP) \cite{vasp}. The projector-augmented wave (PAW) pseudopotential \cite{PAW} and the generalized gradient approximation (GGA) as parameterized by Perdew, Burke, and Ernzerhof (PBE) \cite{PBE} are used in the performence. A Monkhorst-Pack k-point grid of $6\times6\times1$ is adopted to sample the first Brillouin zone. The cutoff energy for the plane wave basis is 350 eV, and the convergence criteria for the self-consistent calcualtion is set to 10$^{-6}$ eV. We do not consider the spin-orbital coupling (SOC) when generating the training-dataset, since it is demonstrated that the SOC has little impact on the calculated atomic force and hence on the subsequent training of the MLFF. The density dependent screened Coulomb (dDsC) dispersion correction is invoked by setting IVDW=4 \cite{ivdw4_1, ivdw4_2}. The twisted MoTe$_2$ moir\'e superlattice is efficiently relaxed by the MLFF and the band structure as well as real space charge density in the equilibrium are calculated by VASP. We use $1\times1\times1$ k-point sampling for supercell calculation. The Chern number can be determined from the eigenvalue at high symmetry points: $e^{i\,2\pi C/3}\!=\!-\xi_{\Gamma_{M}}\xi_{K_{M}}\xi_{K_{M}'}$, where $\xi_{\textbf{k}}$ is the spinful $C_{3}$ eigenvalue at $\textbf{k}$. The $C_{3}$ eigenvalues are obtained using IRVSP package \cite{irvsp}. It is noted that the reference lattice constant in constructing the untwist $3\times3$ and twisted moir\'e superlattice is chosen by fully relaxing the primitive bilayer MoTe$_2$. A vacuum layer of 20 Å alone the $z$ direction is used in these systems to avoid spurious interactions between periodic images.

\end{widetext}



\end{document}